\documentclass[10pt,journal,compsoc]{IEEEtran}

\usepackage{subcaption}
\usepackage{amsmath}
\usepackage{graphicx}
\usepackage{enumerate}
\usepackage{xspace}
\usepackage{xcolor}

\usepackage{dblfloatfix}
\usepackage{amsmath}
\usepackage[ruled]{algorithm}
\usepackage[noend]{algorithmic}

\usepackage{cite,cases,color,array}
\usepackage{algorithmic}
\usepackage{graphicx}
\usepackage{mdwmath}
\usepackage{mdwtab}
\usepackage{eqparbox}
\usepackage{amsopn}
\usepackage{multirow}
\usepackage{graphics}
\usepackage[font=footnotesize]{subfig}
\usepackage{paralist}
\usepackage{xspace}
\usepackage{amsmath}
\usepackage{epsfig}
\usepackage{latexsym}
\usepackage{amsfonts}
\usepackage{amssymb}
\usepackage{comment}
\usepackage{mathrsfs}
\usepackage{cases}

\newcommand{\MYCUT}[1]{{ }}

\def\ie{\textit{i.e.}\xspace}

\def\eg{\textit{e.g.}\xspace}

\newcommand{\eqqref}[1]{Eq.~(\ref{#1})}

\def \ourprotocol{WalkieLokie\xspace}

\allowdisplaybreaks

\begin{document}

\title{WalkieLokie: Relative Positioning for Augmented Reality Using a Dummy Acoustic Speaker}
\author{Wenchao~Huang, Yan~Xiong, Xiang-Yang~Li, Yiqing~Hu, Xufei~Mao, Panlong~Yang  
    \thanks{
    
Wenchao Huang, Yan Xiong, Yiqing Hu are with School of Computer Science and Technology, University of Science and Technology of China.
Email: \{huangwc, yxiong, mfy\}@ustc.edu.cn, huyiqing@mail.ustc.edu.cn.

Xiang-Yang Li is with Department of Computer Science and Technology and TNLIST, Tsinghua University, and Department of Computer Science, Illinois Institute of Technology.
Email: xli@cs.iit.edu.

Xufei Mao is with Department of Software Engineering, and TNLIST, Tsinghua University.
Email: xufei.mao@gmail.com.

Panlong Yang is with Institute of Communication Engineering, PLAUST.
Email: panlongyang@gmail.com.

    }

}

\maketitle

\begin{abstract}
We propose and implement a novel relative positioning system, WalkieLokie, to enable more kinds of Augmented Reality applications, e.g., virtual shopping guide, virtual business card sharing.
WalkieLokie calculates the distance and direction between an inquiring user and the corresponding target.
It only requires a dummy speaker binding to the target and broadcasting inaudible acoustic signals.
Then the user walking around can obtain the position using a smart device.
The key insight is that when a user walks, the distance between the smart device and the speaker changes; and the pattern of \textit{displacement} (variance of distance) corresponds to the relative position.
We use a second-order phase locked loop to track the displacement and further estimate the position.
To enhance the accuracy and robustness of our strategy, we propose a synchronization mechanism to synthesize all estimation results from different timeslots.
We show that the mean error of ranging and direction estimation is 0.63m and 2.46 degrees respectively, which is accurate even in case of virtual business card sharing.
Furthermore, in the shopping mall where the environment is quite severe, we still achieve high accuracy of positioning \textit{one dummy} speaker, and the mean position error is 1.28m.

\end{abstract}

\section{Introduction}

With the rapid development of smart phones and wearable devices, attractive Augmented Reality (AR) apps have been developed, \eg, Sky Map, Wikitude, Augmented Car Finder.
One of AR's key features is to display useful information about a person's surroundings, which relies on location information.
For example, Wikitude uses GPS and inertial sensors to provide interactive information about objects that are seen through the camera of smart devices. 

In this paper, we explore localization techniques to enable more kinds of AR applications on smart devices.
For instance, a person walks in a large shopping mall and a virtual shopping guide recommends the surrounding goods that are new arrivals or on sale; 
or shares her/his virtual business card with people walking around in a party.
Such applications require the knowledge of \textit{relative position} between targets (\eg, goods, person) and inquiring users.

However, current localization systems cannot be directly applied to \textit{relative positioning} satisfactorily due to various limitations.
For instance, these systems can only be used in some places with specified infrastructure being deployed, or require feature-rich hardware serving as target.
More specifically, GPS can calculate the location of outdoor users, but is unavailable in indoor environments.
Pure WiFi-based indoor localization can achieve 3$\sim$4 meter accuracy in absolute positioning and there always exists large errors (\eg, 6$\sim$8m) \cite{2012-MOBICOM-PushlimitWiFi}. 
So the errors are much greater than 4 meters when inferring \textit{relative position} from \textit{absolute positions} of the smartphone and the target. 
Other indoor localization schemes \cite{Xiong:2013:AFI:2482626.2482635,2013-MobiSys-Guoguoenablingfine} are accurate enough (\eg, sub-meter accuracy), but require special-purpose infrastructure or hardware. 
There are schemes calculating relative direction (e.g., Swadloon \cite{DBLP:journals/corr/HuangXLLMYL13}) and distance (e.g., BeepBeep \cite{2007-SenSys-BeepBeephighaccuracy}).
However, Swadloon requires unusual behavior of querying user (\ie, phone-shaking movement) before getting the direction of a target.
More importantly, Swadloon cannot obtain distance from the target.
Though BeepBeep can be added for calculating distance, it requires that the target has rich functions, such as broadcasting and receiving acoustic signals, communication for exchanging data and computation for processing data.
It is feasible when the target uses a smartphone which has all these functions, 
but other applications, such as shopping guide, may prefer a cheaper target device with much fewer functions.


We propose and implement \ourprotocol, which calculates the \textit{relative position} from a user with a smart device to a target for Augmented Reality.
\ourprotocol does not need any infrastructure being deployed,
and the application of \ourprotocol is not limited by places. 
The only requirement of \ourprotocol is that the target is attached with a dummy speaker for broadcasting audio, which can be received by the smart device and then processed to directly infer the relative position.
The dummy speaker merely broadcast audio without requirement of any other features, \eg, audio recording, communication or computation.
Hence, they are widely used and some of them are cheap and simple, such as speaker embedded in user's smart devices, or even loudspeaker originally for sales promotion in a shopping mall.
Moreover, the broadcast audio is inaudible that the loudspeaker, which used to be a noisy tool for sales promotion, can now be ``silent'' for the same job by ``broadcasting'' its relative position.

Our work is based on the observation that when a user walks, the distance between the object and the user changes; and the pattern of \textit{displacement} (variance of distance) relates to the relative position.
In other words, by letting a device receive and analyze the signal (audio signal mixed with non-audible signal) 
broadcast by a target (speaker), we are able to track the displacement and further compute the relative position 
 accurately and efficiently, i.e., finishing both ranging and direction estimation at the same time.

However, we have to solve a number of issues in our scheme.
First, since the displacement is relatively small, the practical challenge is how 
 to obtain the position precisely when a user walks for only very few steps.
Second, when the user is far from the speaker, the measured displacement is 
 prone to be influenced by noises and the difference of displacement becomes more indistinguishable such that 
 even a tiny error in the measurement could cause large errors in positioning.
Hence, both accurate displacement measurement scheme and robust positioning strategy
 are needed.


In order to get an accurate displacement measurement scheme, we track the phase of 
 the signal (corresponding to the displacement) utilizing the second-order Phase Locked Loop (PLL), 
 which could avoid jitters and has high accuracy when the signal is weak.
Hence, the distance and direction could be computed accurately when a user is close to
 a speaker.
Next, as we have mentioned above, the estimated position may have a bigger error
 when a user is far from a speaker. 
In this case, we adopt two strategies to further improve the accuracy and robustness.
One is to utilize the measurement results when the user is close to the speaker if
 available.
Otherwise, we synthesize all the estimations (longer path a user passed) by the synchronizing scheme.
The main idea is based on the following observation:
Since the distance can be obtained according to the difference between the sent time and received time of a signal,
where the receiving time is directly computed but the sent time is unknown for the receiver, we add a periodical pulse into the audio to get the sent time in a novel way.
Specifically, when a good estimation is obtained, the distance along with the sent time of the pulse is calculated.
Hence, the sent time of the later periodical pulses is predicted, which can infer the distance according to sent time and receiving time of the pulses.


\ourprotocol also addresses a number of practical issues based on the main solution:\\
\textbf{The user frequently turns the walking direction:} we provide enhanced algorithm that gathers all the pieces of small linear segments at different directions to obtain the position.\\
\textbf{Multipath effects on pulse detection:} \ourprotocol detects arrival time of all the pulses, including the pulse directly from sender and also reflected ones.
Then it eliminates the false pulses by leveraging the result of PLL.\\
\textbf{Non-Line-of-Sight (NLoS):} \ourprotocol uses historical position results, and infers the current position by additionally using inertial sensors; once the smart device is within the coverage of the signal, \ourprotocol updates accurate position by synchronization.\\
\textbf{Device diversity:} The main problem is serious clock drift of normal dummy speaker, otherwise the receiver obtains wrong receiving time of periodical pulses and further wrong distance in synchronization.
We leverage the result of PLL, and calibrate the clock precisely in case that the receiver is static for only a few seconds.
\\
\textbf{Device Orientation:} Different orientation of the speaker or receiver affects the quality of the received signal.
We find that the quality mainly affects the result of PLL and further the displacement. 
More specifically, when the signal quality is poor for certain orientation, the tracked displacement becomes smaller than the real displacement.
To enhance the accuracy, we make calibrations on tracked displacement based on our measurements.
\\
\textbf{Conflicts of multiple signals:} In \ourprotocol, the periodical pulses in synchronization possess bandwidth, which limits the number of co-existing signals.
We carefully design the pulse that possesses narrow bandwidth, and also show the way of supporting more number of co-existing speakers.
\\
\textbf{Noisy environment:} \ourprotocol uses Band Pass Filter to eliminate the noises and it works well in the noisy shopping mall.

We implement \ourprotocol and evaluate the performance of all the components separately with 
 several types of cases and then the performance of \ourprotocol.
 \\\textbf{a).} For the case when a user is within 8 meters away from a speaker, the mean error of ranging and direction estimation is $0.63m$ and $2.45^o$.
It shows considerable accuracy when a user shares virtual business card with surroundings.
 \\\textbf{b).} When the user is within $20m$ and uses synchronization for positioning, the ranging and direction estimation errors are less than $0.32m$ and $2.81^o$ at the percentage of $80\%$ respectively.
 Note that the results in this case only infer the accuracy of the subcomponent (synchronization), instead of the total accuracy of \ourprotocol. 
 \\\textbf{c).} We combine all the work together and evaluate \ourprotocol in a severe environment, \ie, the noisy shopping mall.
 We conduct the experiment in 2 cases: 
 \begin{itemize}
     \item \textbf{c1).} \textit{Relative} positioning of \textit{one} speaker;
 \item \textbf{c2).} \textit{Absolute} positioning using \textit{multiple} speakers (\ie, ordinary indoor localization). 
 \end{itemize}
We put 5 dummy speakers in a $600m^2$ area, and the positions of speakers are limited to be deployed, (just at the side of aisle, instead of the position on the ceiling). 
Even in this case, by using only \textit{one} of these speakers, \ourprotocol also achieves the mean error of $1.28m$ for the \textit{relative} position.
Hence, a user knows the accurate relative position of a virtual shopping guide attached with a dummy speaker.

To explore the possibility of \textit{absolute} positioning, we also use all the 5 speakers as anchors for localization and the mean error is $0.89m$, where each position is covered by the signal of less than 2 speakers on average.
Since normal anchor-based systems only obtain distance or direction but cannot calculate both metrics, they require at least 3 anchors for trilateration. 
Hence it shows another advantage of \ourprotocol that it is robust when the anchors are sparse in deployment.

The rest of the paper is organized as follows.
We first present the overview of \ourprotocol in Section \ref{sec:overview}.
We propose the position estimation 
 based on displacement tracking in Section \ref{sec:posest} and 
 displacement tracking method in Section \ref{sec:pll}.
We give the details of the synchronization in Section \ref{sec:sync}.
We report our extensive experimental results in Section \ref{sec:exp}.
We review some related work in Section \ref{sec:relatedwork}.
We conclude the paper in Section \ref{sec:conclusion}.

\section{Overview}
\label{sec:overview}

\subsection{Problem Description}
\ourprotocol calculates \textit{relative position} between a user with a smart device and a target attached with a dummy speaker,
where the relative position can decompose into distance and direction from the smart device to the dummy speaker.
The dummy speaker merely broadcasts inaudible audio without the requirement of any other features.
The smart device has a microphone and inertial sensors (\ie, compass, accelerometer, gyroscope), which are common components in almost all smart devices.

\subsection{Intuitive Solution}
\label{sec:subsolution}
The key insight of our paper is that when a user walks along a line, the pattern of displacements
 from the user to a dummy speaker is related to relative position directly.

\begin{figure}[t]
\begin{subfigure}[h]{0.14\textwidth}
    \begin{center}
        \includegraphics[width=1.0in]{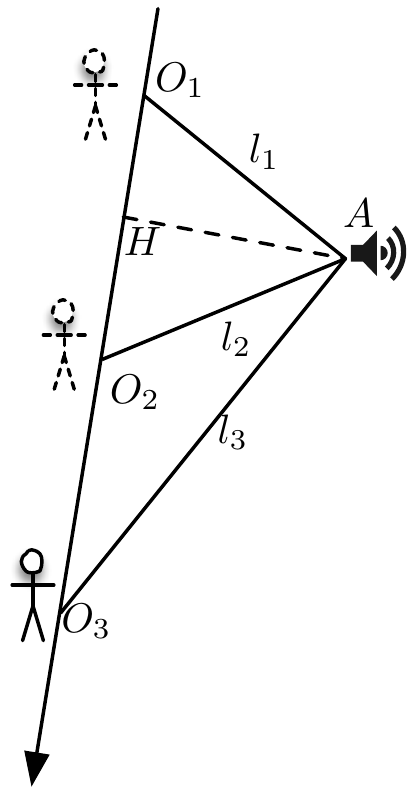}
    \end{center}
    \caption{Brief Example. }
    \label{fig:example}
\end{subfigure}
\begin{subfigure}[h]{0.24\textwidth}
    \begin{center}
        \includegraphics[width=2.6in]{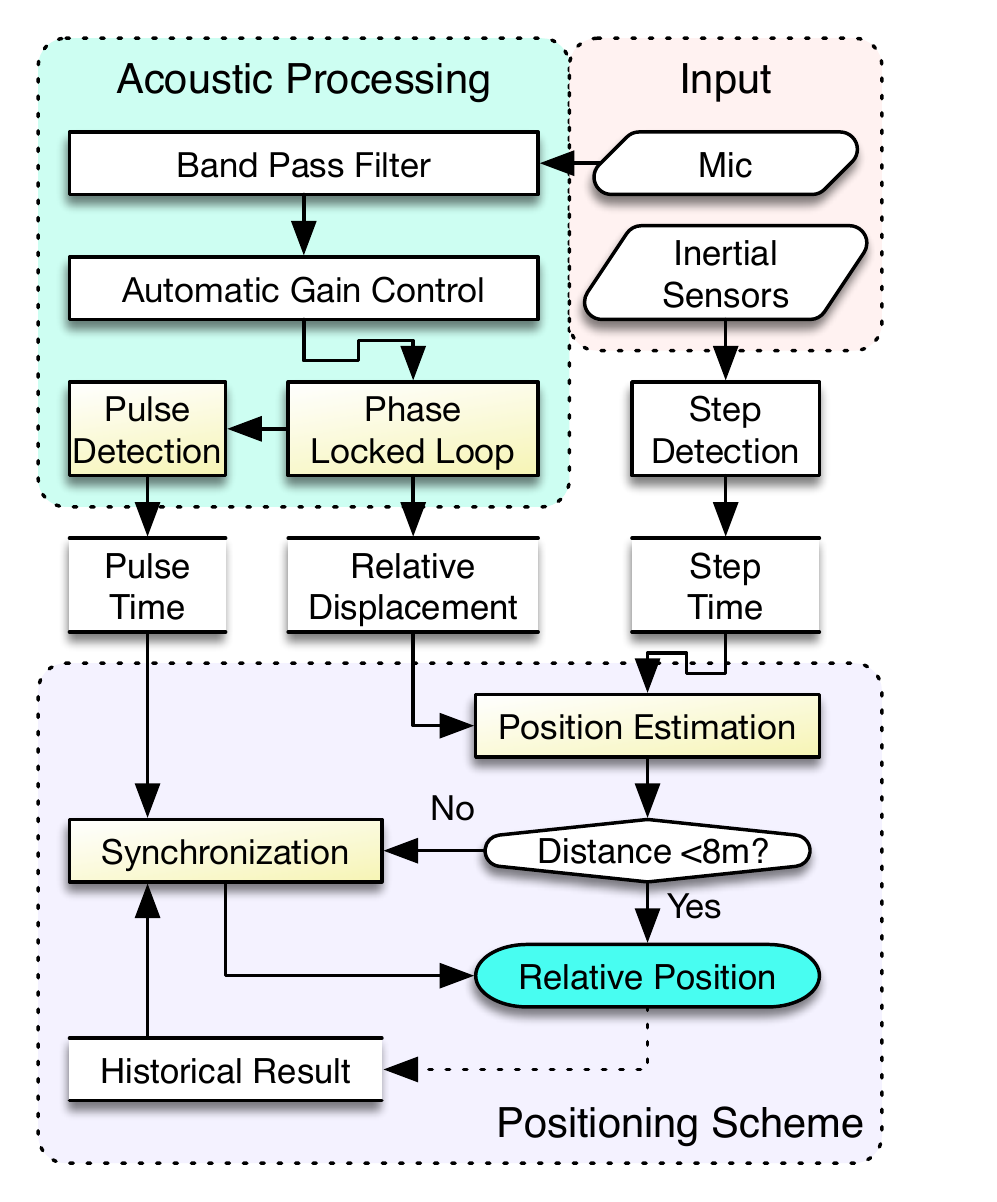}
    \end{center}
    \vspace{-0.1in}
    \caption{Architecture of \ourprotocol.}
    \label{fig:overview}
\end{subfigure}
    \caption{Example of relations between displacement and relative positions, and architecture of \ourprotocol.}
    \vspace{-0.2in}
\end{figure}

We illustrate the intuitive solution on a simple case in Figure \ref{fig:example}, where a user walks and steps at $O_1$, $O_2$, and $O_3$.
Suppose the displacements $d_1$(=$l_1-l_2$) and $d_2$(=$l_2-l_3$) are measured beforehand
 and the user's stride length ($|\overline{O_1O_2}|$) is given.
Intuitively, $d_1\approx0$ infers that $O_1$ and $O_2$ are close to $H$ where $\overline{AH} \bot \overline{O_1O_2}$ and $d_2<0$,
 which tells us that the speaker is at the back of the walking user,
 hence infers the coarse-grained direction.
Another observation is that when the distance $|\overline{AH}|$ increases, the value of 
 $|d_2-d_1|$ decreases, which infers the coarse-grained distance as well.
So, the relative position between $O_1$ and $A$ can be estimated.

\subsection{Main Technical Issues}
From the above example, the following main technical issues required to be solved:

\noindent \textbf{Formal solution of relative positioning (Section \ref{sec:posest}).} Given the real-time relative displacement, we need to calculate the precise relative position, instead of coarse-grained one. 

\noindent \textbf{Tracking relative displacement (Section \ref{sec:pll}).} The relative displacement needs to be tracked before relative positioning. 
Note that to the best of our knowledge, current approaches cannot directly obtain distances $l_1$ without synchronization between the receiver and the dummy speaker.
They require additional capabilities of the speaker, such as communication capability that exchanges synchronization information \cite{2007-SenSys-BeepBeephighaccuracy}.
Instead, we calculate the displacement: $d_1 (=l_1-l_2)$ by PLL.

\noindent \textbf{Extended solution when distance is longer (Section \ref{sec:sync}.}
When $|\overline{AH}|$ becomes much longer, $|d_2-d_1|$ is much smaller.
Since there are errors on tracking the displacement $d_1$, $d_2$, the ideal case is that small change of distance corresponds to large value of $|d_2-d_1|$, which results in high accuracy of calculated distance.
However, when $|\overline{AH}|$ becomes much larger, it is the opposite case that tiny error on measuring $d_1$ or $d_2$ will result in large error on calculating $|\overline{AH}|$. 
Hence, the accuracy of ranging declines when $|\overline{AH}|$ becomes larger, and we need extended solution in this case.
Note that the accuracy of direction finding is not much affected that we mainly propose the extended solution for ranging.


\subsection{Architecture}
To solve the technical issues, we divide \ourprotocol into 3 main components in Figure \ref{fig:overview}: input of smart device, acoustic processing, and positioning scheme. 

\noindent \textbf{Input:} The microphone and inertial sensors are used in \ourprotocol. 
The microphone records audio for acoustic processing.
The inertial sensors mainly serve as a step counter, which records the time when the user steps on the ground.
When the user turns direction, the angle of user's rotation is also calculated by the gyroscope.

\noindent \textbf{Acoustic processing:} This component generates intermediate results preparing for the positioning scheme.

One result is \textit{relative displacement}, which is tracked by analyzing the recorded audio (in Section \ref{sec:pll}). 
The audio firstly passes through the Band Pass Filter (BPF) that the signal at the specified frequency passes and other signals including human voice and noises are eliminated. 
Then, the filtered signal is processed by Automatic Gain Control (AGC), and then the amplitude of the signal is close to constant.
The signal then passes through our carefully-designed Phase Locked Loop (PLL), and the phase of the signal, which is proportional to relative displacement, is tracked.

Another intermediate result provides additional information for the extended solution.
More specifically, we encode periodical pulses in the sent signal, and the smart device detects the corresponding pulses to determine the \textit{receiving time of the pulses} (in Section \ref{sec:sync}). 
The problem is that the pulse should take very little bandwidth, otherwise the number of concurrent speakers is much limited.
We carefully encode the signal to solve this problem, and design the pulse detection algorithm to precisely determine receiving time of the pulses.
Note that the pulse detection analyzes tracked phase from output of PLL, for we directly modulate the phase, rather than the raw audio, to encode the pulses in order to save bandwidth.

\noindent \textbf{Positioning scheme:} The scheme calculates position by receiving the intermediate results.
The scheme firstly estimates position by using the relative displacements and user's step time.
It leverages the intuitive solution in Section \ref{sec:subsolution}, which is formally illustrated in Section \ref{sec:posest}. 
Then, if the computed distance is very short ($<8m$), the calculated position is accurate enough and accepted as valid result.
Otherwise, the calculated direction is accurate, but the calculated distance is inaccurate.
In this case, the scheme invokes synchronization in Section \ref{sec:sync} to compute the relative position.
The synchronization uses the historical results of relative position to infer the distance.
By additionally using the historical receiving time of the pulses, the sending time of the periodical pulses is then calculated.
The accurate distance is then inferred from the detected receiving time and the predicted sending time of the current pulse.

\section{Estimation of Relative Position}
\label{sec:posest}

In this section, we propose the method on distance and direction estimation from smart device to speaker, \ie, the relative position.
The intuition is that when the user walks, there is a unique pattern of displacement according to relative position.
Hence, we use the displacements (calculated in section \ref{sec:pll}) to deduce user's positions. 

\begin{figure}[htpb]
    \centering
        \begin{center}
            \includegraphics[height=1.3in]{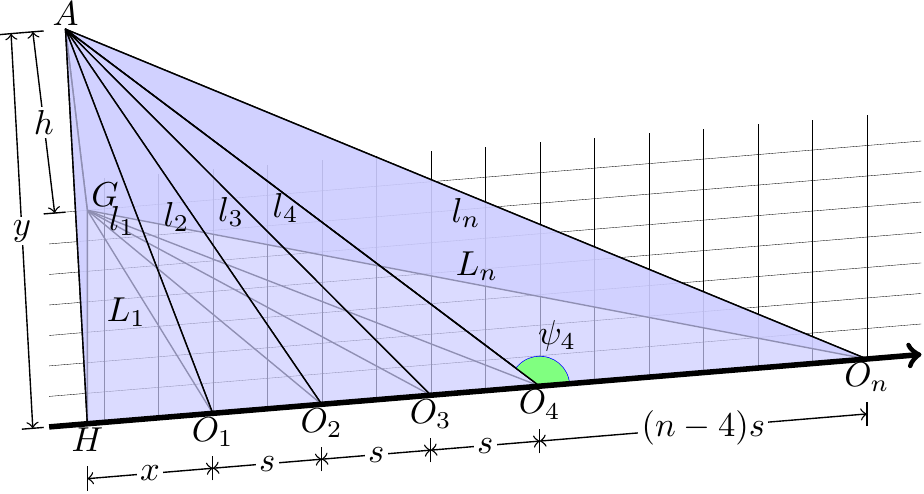}
        \end{center}
        \caption{Positioning when the user walks along a line.}
        \label{fig:distance}
    \vspace{-0.2in}
\end{figure}

\subsection{Positioning When User Walks along a Line}
To estimate the distance, we first consider a simple scenario when a user starts walking from $O_1$ and steps at $O_2$, $O_3, \dots, O_n$, shown in Figure \ref{fig:distance}.
$A$ is the position of the speaker.
Denote the height of the speaker relative to the smart device as $h=|\overline{AG}|$. 
Assume the stride length is close to constant $s=\overline{O_iO_{i+1}}$.
Both $h$ and $s$ are assumed to be given and used for distance and direction estimation.
The other inputs are the displacements of all the steps, \ie, $d_i=l_i-l_{i+1}$ for the step $\overline{O_iO_{i+1}}$, which are calculated using PLL.
Observe that the distance from the speaker to $\overline{O_iO_{i+1}}$ is constant $y=|\overline{AH}|$, where $\overline{AH} \perp \overline{O_1O_n}$.
Hence, we first estimate $x=|\overline{HO_1}|$ and $y$ from those inputs and then estimate the position at each step point $O_i$ according to $x$ and $y$.



Intuitively, $x$ and $y$ can be found by traversing the positions and using maximum likelihood estimation.
Specifically, as $|HO_i|= x+(i-1)s$, $i=1,2,3,\dots$, denoting that
   \begin{eqnarray}
\label{eq:mine0}
l_i'=&\sqrt{y^2+(x+(i-1)s)^2}\\
e_i=&l_i'-l_{i+1}'-d_i
\label{eq:mine}
    \end{eqnarray}
Then $e_i=0$ if $d_i$ is accurate.
Hence, for $n$ displacements $d_1, d_2, \dots, d_n$, $x$ and $y$ can be solved from above $n$ equations by $(x,y)=\underset{x,y}{\arg\min}\sum_{i=1}^ne_i^2$.
Here we use the Newton's Method \cite{Madsen04immmethods} to reduce the computation overhead.

%
%

Observe that $L_i=|\overline{GO_i}|$ and $\psi_i'=\angle GO_iO_n$, instead of $l_i$ and $\psi_i$, are the horizontal distance and direction and used for positioning when $x$ and $y$ are estimated, we make the distance and direction results in the following equations:

\begin{equation}
    \begin{cases}
        L_i=&\sqrt{(x+(i-1)s)^2 + y^2-h^2} \\
        \cos\psi_i'=&  - \frac{x+(i-1)s}{L_i} 
    \end{cases}
    \label{eq:estposition}
\end{equation}

\subsection{Synthesizing When User Turns Direction}

\begin{figure}[tpb]
    \centering
    \includegraphics[width=2.7in]{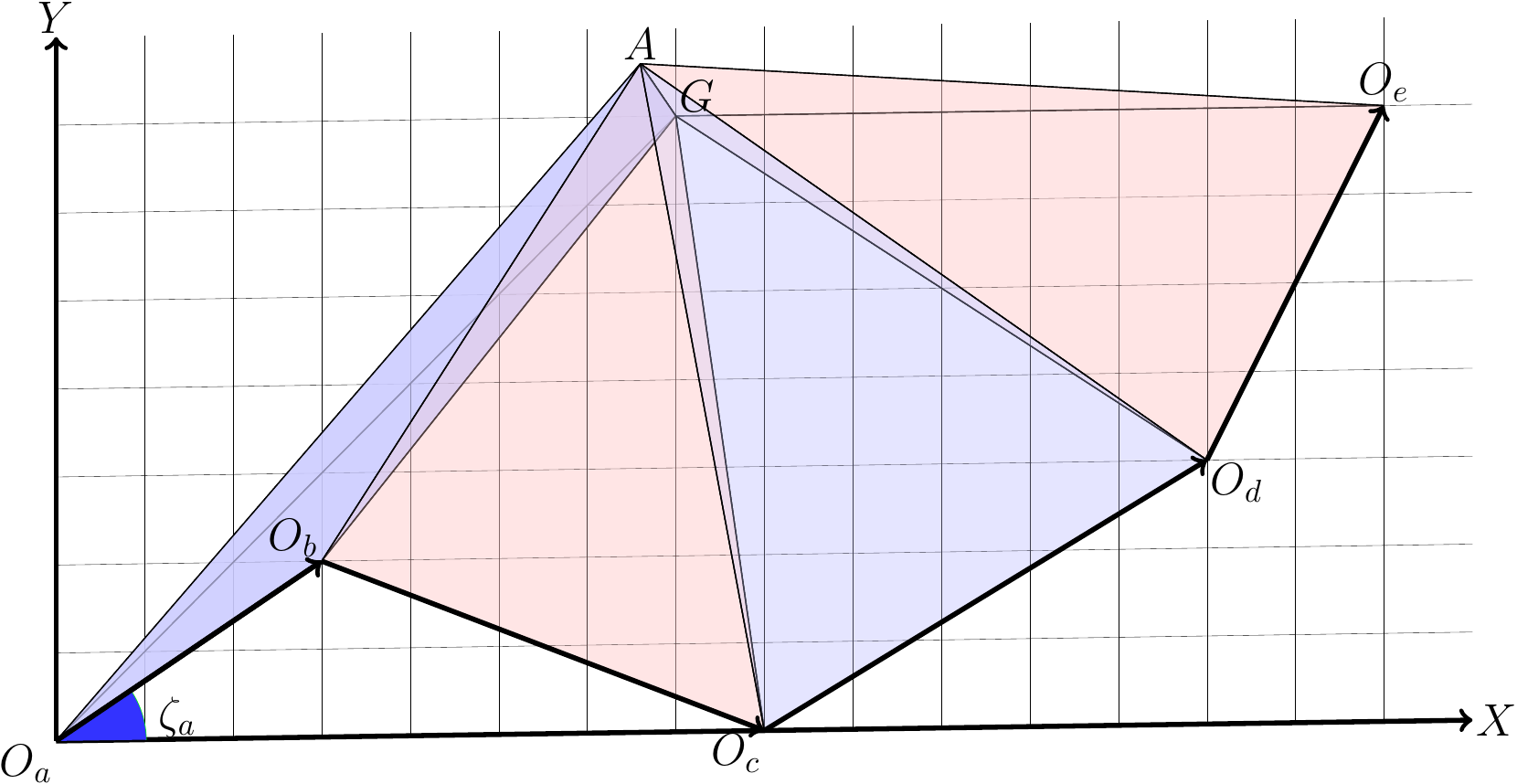}
    \caption{Positioning when the user walks and turns.}
    \label{fig:posuserturns}
    \vspace{-0.2in}
\end{figure}

When a user turns direction while walking, we can always calculate the relative position as follows.
Assume that the user starts from $O_a$ and walks along the linear segment $\overline{O_aO_b}$, $\overline{O_bO_c}$, $\overline{O_cO_d}$, $\overline{O_dO_e}$ in Figure \ref{fig:posuserturns}.
We use the calculated displacements in this case.
We also use the step counter to estimate the linear length $n_as$, $n_bs$, $n_cs$, where $s$ is the stride length and  $n_a$ is the number of steps when user walks from $O_a$ to $O_b$.

\noindent \textbf{Calculating angle of turning direction:} We calculate angle of user's rotation mainly by using gyroscope, when the user turns direction. 
Though Zee \cite{2012-MOBICOM-Zeezeroeffort} can directly calculate walking direction, it is mainly affected by inaccurate compass and usually cannot distinguish whether the user is walking forward or backward along a direction. 
\ourprotocol does not require the knowledge of absolute direction of user's walking for we only need to know relative position from the user to the target. 
It only requires the angle of user's rotation when the user turns, which is used for calculating position in this section.
For instance, assume the initial walking direction is $\zeta_a$ and the following direction is $\zeta_b$. We do not calculate $\zeta_a$ or $\zeta_b$ by magnetic sensor, but directly calculate the difference of walking direction, \ie, $\zeta_b-\zeta_a$, from the gyroscope.
The purpose is to avoid errors caused by magnetic sensor of the smart device, where the errors might be huge in indoor environments.
Note that  $\zeta_a$ can be eliminated as we will convert the position in WCS (World Coordinate System) into the one in RCS (relative coordinate system), as mentioned in section \ref{sec:exp}.
Hence, our problem is much easier that the gyroscope can accurately calculate the angle of user's rotation.
By using this rotation angle together with the step detection, we can get the walking trace without knowing the relative position.
Furthermore, when the relative position is obtained by additionally using the acoustic signal, the coarse-grained position can be obtained according to the same technique by only using the inertial sensors, if there is no signal received (\ie, NLoS effects).

\noindent \textbf{Calculating position:} Now we calculate relative position $G(g_x,g_y)$, which is the projection of acoustic speaker $A$ at a horizontal plane and is at the same height with the receiver.
Denote $O_a$ is at $(0,0)$, we estimate the next positions $O_c, O_d\dots$ from the step counter and gyroscope.
For example, $O_c$ is at the position $(c_x,c_y)=(n_as\cos(\zeta_a)+n_bs\cos(\zeta_b), n_as\sin(\zeta_a)+n_bs\sin(\zeta_b))$, and so forth.
Given the calculated displacement $d_{c_1},d_{c_2}, \dots, d_{c_{n_c}}$, similar to \eqqref{eq:mine0}, \eqref{eq:mine}, the distance from each stride point to $G$ is 
\begin{equation}
    l_{c_i}=\sqrt{[c_x+(i-1)s\cos(\zeta_c)]^2+[c_y+(i-1)s\sin(\zeta_c)]^2+h^2}
    \label{eq:userturns}
\end{equation}
Denote the calculated error at the $i$th step  along line $\overline{O_cO_d}$ is 
\begin{equation}
    e_{c,i}=l_{c_i}-l_{c_{i+1}}-d_{c_{i}}
\end{equation}
Hence, we obtain the position of $G$ using the following equation:
\begin{equation}
(g_x,g_y)=\underset{g_x,g_y}{\arg\min}\sum_{i\in\{a,b,c,d,e\}}\sum_{j=1}^{n_c-1}e_{i,j}^2
\end{equation}

\begin{figure*}[htpb]
    \begin{subfigure}[b]{0.33\textwidth}
        \begin{center}
            \includegraphics[width=2.1in]{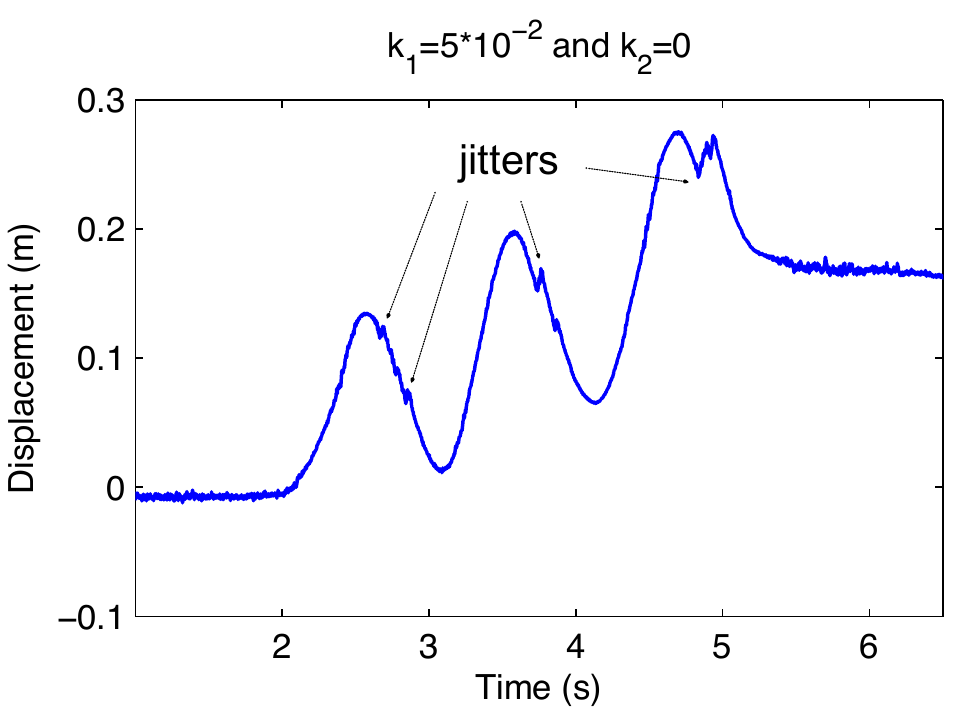}
        \end{center}
        \caption{First order, large $k_1$}
        \label{fig:plla}
    \end{subfigure}
    \begin{subfigure}[b]{0.33\textwidth}
        \begin{center}
            \includegraphics[width=2.1in]{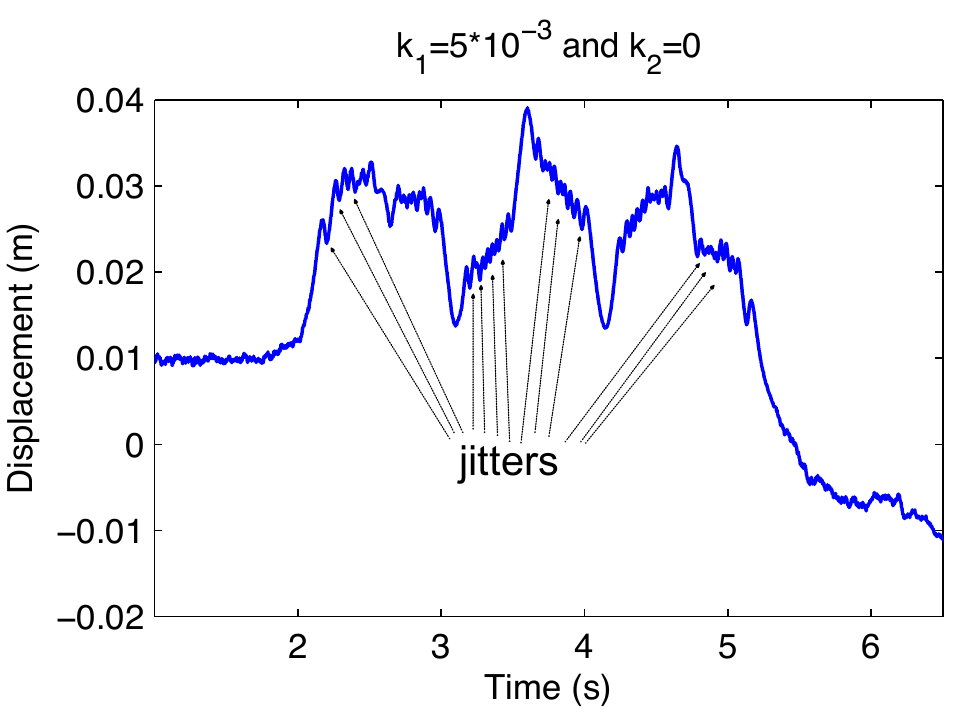}
        \end{center}
        \caption{First order, small $k_1$}
        \label{fig:pllb}
    \end{subfigure}
    \begin{subfigure}[b]{0.33\textwidth}
        \begin{center}
            \includegraphics[width=2.1in]{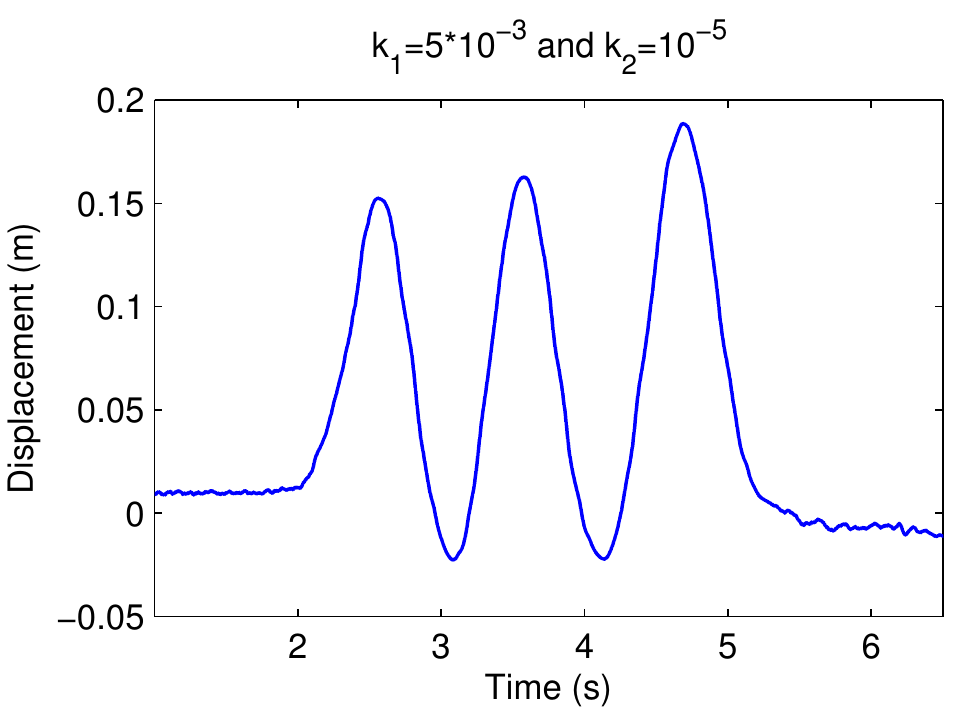}
        \end{center}
        \caption{Second order}
        \label{fig:pllc}
    \end{subfigure}
    \caption{Calculated displacement by PLL with different orders and parameters, when the signal is weak. }
    \label{fig:displacement}
    \vspace{-0.2in}
\end{figure*}

\section{Tracking Displacement}
\label{sec:pll}
In this section, we show how we design the acoustic wave emitted by the dummy speaker and infer the displacement from the acoustic wave when the user walks.


\subsection{Brief Design of the Acoustic Wave}
The modulated wave in audio is used for two purposes: displacement tracking and synchronization.
Hence, the wave $s(t)$ contains two parts respectively.
More specifically, we formally define the wave in the following equations,
\begin{equation}
s(t)=\begin{cases}
s_{1}(t) & kT_2\leq t<kT_2+T_{1}\\
s_{2}(t) & kT_2+T_{1}\leq t<(k+1)T_2
\end{cases}
\end{equation}
where $T_2=0.25s$ is the cycle of the wave and $k$ is the natural number. 
$T_1=0.16s$ is the duration of $s_1(t)$ in each cycle. 

We mainly use $s_1(t)$ for tracking the displacement. 
Intuitively, $s_1(t)$ is a sine wave, and phase of the corresponding received signal $r_1(t)$ is changed when the distance changes.
We prove in the following subsection that the phase of $r_1(t)$ is proportional to relative displacement.
So by tracking the phase of $r_1(t)$, the displacement is tracked.
The displacement tracking algorithm (PLL) will be detailed in this section. 
The relation between phase and displacement is illustrated in section \ref{sec:placedis} and we explain how to preprocess signal and track phase in section \ref{sec:preprocess} and \ref{sec:subpll} respectively.

Note that $s_2(t)$ is not only used for synchronization, but also capable of displacement tracking, like $s_1(t)$.
As a result, the measurement of displacement is rarely affected by additional function of synchronization.
In section \ref{sec:sync}, we will discuss the use of $s_2(t)$.

\subsection{Phase \& Displacement}
\label{sec:placedis}
In order to track displacement, we define $s_1(t)$ as follows:  
\begin{equation}
s_1(t)=\cos(2\pi ft)
    \label{eq:s1}
\end{equation}
where $f$ is the frequency. To make the audio inaudible and to have the frequency supported by commercial speaker, we set $17000Hz <f< 24000Hz$.

On receiving the signal $r_1(t)$, there is a phase shift $\phi$ compared with $s_1(t)$, such that $r_1(t)=\cos(2\pi ft+\phi)$.
For instance in Figure \ref{fig:example}, the displacement \cite{DBLP:journals/corr/HuangXLLMYL13} is 
\begin{equation}
    d=l_1-l_2=\frac{v_a}{2\pi f}(\phi_2-\phi_1)
    \label{eq:relationphi}
\end{equation}
where $\phi_1$ and $\phi_2$ is the calculated phase at $O_1$ and $O_2$ respectively and $v_a$ is the travelling speed of acoustic wave.

\subsection{Preprocessing Received Signal}
\label{sec:preprocess}

Before tracking phase $\phi$ from $r_1(t)$, we have to preprocess the received signal.
For the sent signal $s_1(t)$, the actual received signal $r_{raw}(t)$ does not equal to $r_1(t)$. 
Its amplitude $A(t)$ always changes and it is also mixed with noises $\sigma(t)$.
We denote $r_{raw}(t)=A(t)\cos(2\pi ft+\phi(t))+\sigma(t)$.
Hence, we need to firstly eliminate $A(t)$ and $\sigma(t)$ before tracking the phase $\phi(t)$.

To eliminate the noise $\sigma(t)$, we let $r_{raw}(t)$ pass through a Band Pass Filter (BPF).
The processed signal $r_{filter} \approx A(t)\cos(2\pi ft+\phi(t))$.
$r_{filter}$ is then processed by Automatic Gain Control (AGC) \cite{rice2008digital}. 
After that, $A(t)$ is removed and the signal can be seen as $r_1(t)$ \cite{DBLP:journals/corr/HuangXLLMYL13}.

\subsection{Tracking the Phase}
\label{sec:subpll}
To track phase for inferring displacement, we adopt the second-order Phase Locked Loop (PLL) to track the phase when the smart device moves, rather than the ordinary first-order PLL.
PLL is a classical method in signal processing and can be regarded as a device that tracks the phase and frequency of a sinusoid. 
In our design, it is implemented purely by software due to the limited capabilities of smartphone platform. 

\begin{figure}[htpb]
    \begin{center}
        \includegraphics[width=2.8in]{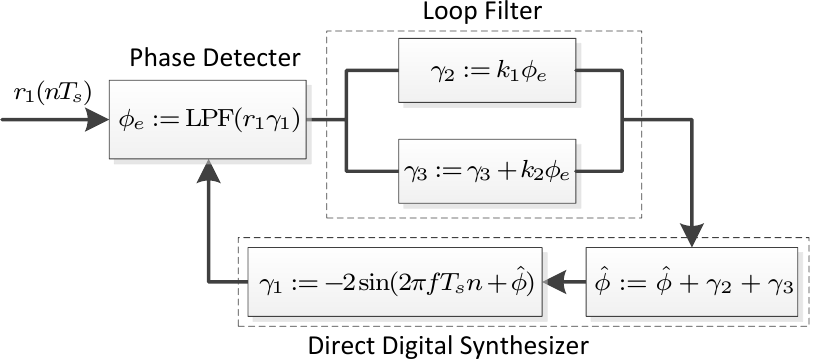}
    \end{center}
    \vspace{-0.1in}
    \caption{Design of the Second-Order Phase Locked Loop. }
    \label{fig:pll}
    \vspace{-0.1in}
\end{figure}

\begin{figure*}[t]
    \begin{subfigure}[b]{0.245\textwidth}
        \begin{center}
            \includegraphics[width=1.8in]{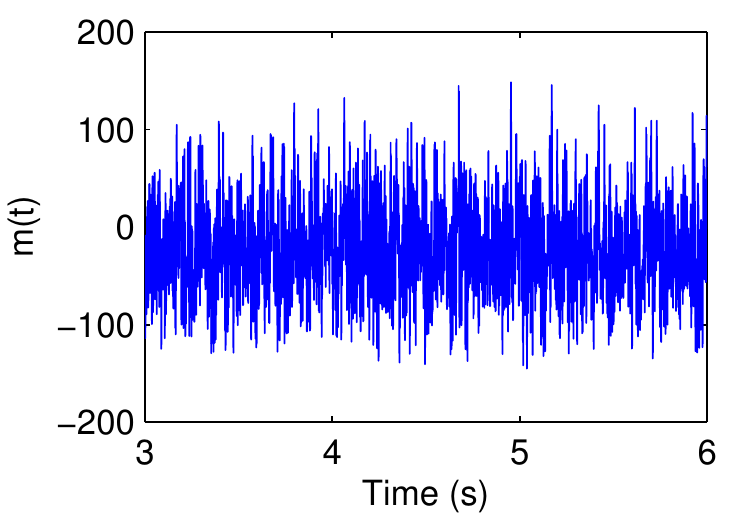}
        \end{center}
        \caption{m(t) of Weak Signal}
        \label{fig:fmtbad}
    \end{subfigure}
    \begin{subfigure}[b]{0.245\textwidth}
        \begin{center}
            \includegraphics[width=1.8in]{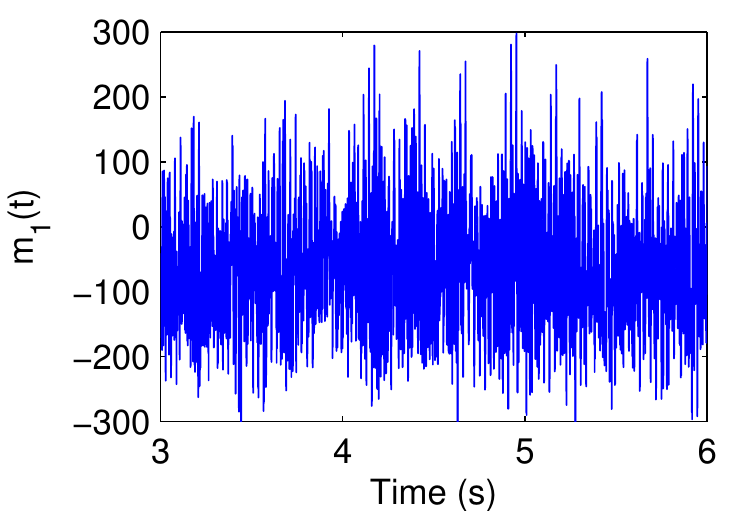}
        \end{center}
        \caption{$m_1(t)$ of Weak Signal}
        \label{fig:fmtbad1}
    \end{subfigure}
    \begin{subfigure}[b]{0.245\textwidth}
        \begin{center}
            \includegraphics[width=1.8in]{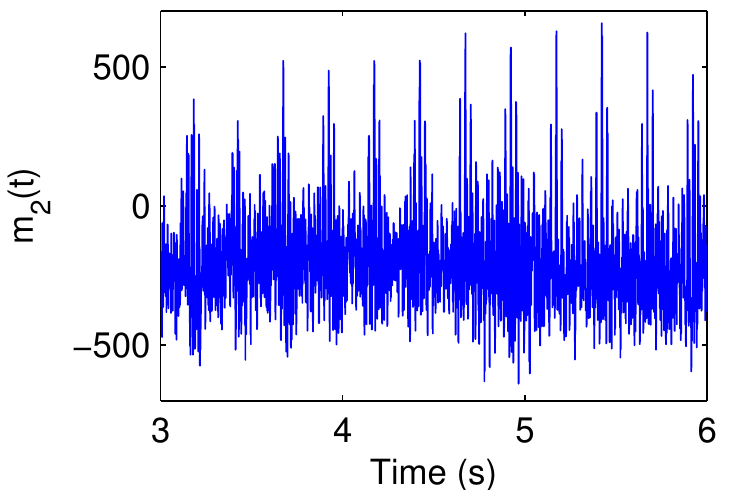}
        \end{center}
        \caption{$m_2(t)$ of Weak Signal}
        \label{fig:fmtbad2}
    \end{subfigure}
    \begin{subfigure}[b]{0.245\textwidth}
        \begin{center}
            \includegraphics[width=1.8in]{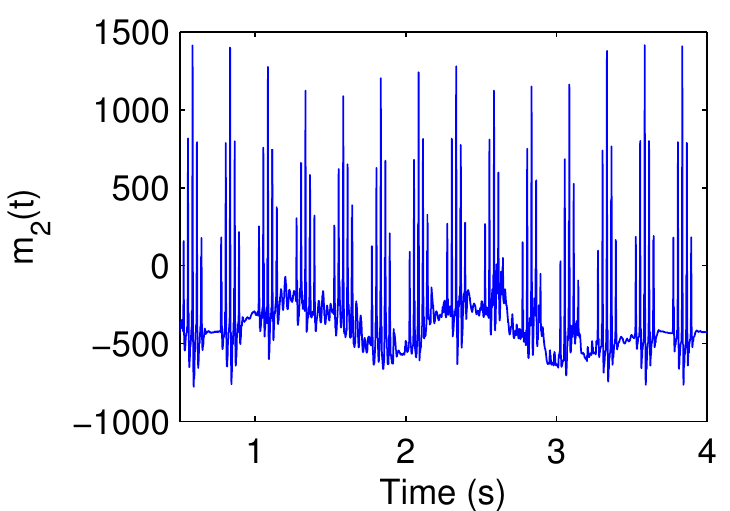}
        \end{center}
        \caption{$m_2(t)$ of Good Signal}
        \label{fig:fmtgood2}
    \end{subfigure}
    \caption{Detection of the arrival time of the pulse. }
    \label{fig:syncgood}
    \vspace{-0.2in}
\end{figure*}

We show our design of PLL in Figure \ref{fig:pll}.
The PLL contains three main components: phase detector, loop filter and direct digital synthesizer (DDS).
The phase detector detects the difference $\phi_e=\phi-\hat \phi$, where $\hat \phi$ is the estimation of $\phi$.
According to $\phi_e$, the loop filter analyzes and predicts the offset $\gamma_2+\gamma_3$ of $\hat \phi$ for the next cycle of the loop, where the variance of $\gamma_2, \gamma_3$ is affected by parameter $k_1$ and $k_2$ respectively.
The DDS updates the next $\hat \phi$ by adding the offset and prepares $\gamma_1$ for the next phase detection.

In the process of the phase detector in Figure \ref{fig:pll}, for the $n$th input $r_1(nT_s)$, $r_1 \gamma_1= \sin(\phi-\hat \phi) -\sin(4\pi f T_s n +\phi + \hat \phi)$. 
Here we denote $T_s$ as sampling period of received signal.
As $\phi_e=\textrm{LPF}(r_1 \gamma_1)$ where LPF is the low pass filter, the high frequency component of $r_1\gamma_1$ is eliminated and $\phi_e\approx \sin(\phi - \hat \phi)$.
If the phase is locked (\ie, $\hat \phi$ is close to $\phi$), $\sin (\phi- \hat \phi) \approx \phi- \hat \phi $. 
Hence $\phi_e \approx \phi - \hat \phi$.


In Figure \ref{fig:pll}, the Loop Filter is the key part of PLL. There have been many proposals on design of loop filter \cite{best2003phase}, 
and the type and parameter of Loop Filter should be carefully chosen for different purposes. 
Here we adopt a second-order filter, \ie, the proportional-plus-integrator \cite{rice2008digital} filter, as the Loop filter.
It uses two updated variables $\gamma_2$, $\gamma_3$ and two constant parameters $k_1$, $k_2$.
Particularly, if $k_2=0$, it degrades to a first-order PLL.

We explain why the first-order PLL cannot be used in our case.
When the phone is static and the PLL becomes stable after several cyclic loops, $\phi_e\approx 0$ and $\hat \phi$ is close to constant.
Hence in Figure \ref{fig:pll}, $\gamma_2 \approx 0$ and $\gamma_2+\gamma_3 \approx 0$ which infers $\gamma_3 \approx 0$.
It means that $k_2$ can be eliminated and the first-order PLL is sufficient.
However, if the phone moves, and we still use the first-order PLL, the performance is good in case of high signal-to-noise ratio (SNR) but also limited by SNR. 
For instance, assume the user moves at a constant relative speed and $\phi$ increases $\Delta \phi$ per $T_s$, \ie, the cyclical time of the loop.  
When the PLL becomes close to stable,  $\phi_e \approx \Delta \phi +\phi_s$ where $\phi_s$ is error caused by random noises.
If the SNR is high that $\phi_s \ll \Delta \phi $, we can set $k_1>\Delta\phi$ to let $\hat \phi$ catch up with the variation of $\phi$.
However, when increasing the value of $k_1$, as the magnitude of $\gamma_2$ increases, $\hat \phi$ becomes unstable and tends to be affected by noises.
The bad case is that the phase, which is actually $\phi$, is intended to be locked or already locked to $\phi+2\pi$ or $\phi-2\pi$.
We call this phenomenon the \textit{jitter} for convenience.
The error of the corresponding displacement is $\frac{v_a}{2\pi f}2\pi \approx 1.8cm$ which affects the accuracy of position estimation in section \ref{sec:posest}.

To show the limitation in the experiment, we let the user hold the phone for a while, move the phone forward to the speaker and backward for three times, and finally stop at the starting point.
In Figure \ref{fig:displacement}, we show the result of PLL with different parameters, when the acoustic signal is weak, \ie, $l=32m$. 
In Figure \ref{fig:plla}, the $k_1$ is large enough to catch up with the real displacement.
However, it is affected by the noises and sometimes cannot lock when moving. 
It results in occasional jitters on the up-and-down curve.
Then, the calculated displacement from the start to the end, which should be close to 0, accumulates to $17cm$ after total moving length of about $100cm$.
On the contrary in Figure $\ref{fig:pllb}$ where $k_1$ is small, the calculated phase displacement cannot catch up with the real phase and jitters frequently when moving.
Hence, there is limitation of using first-order PLL for supporting both high-speeding moving and high noises.

Therefore, for solving the above problem, the updated component $\gamma_3$ is added, which turns the first-order PLL into the second-order one.
$\gamma_3$ can be seen as the phase variation $\Delta \phi$ per $T_s$, which corresponds to the relative speed from the phone to the speaker.
If the PLL becomes stable, in each cyclic loop, the loop filter predicts next phase with the added $\gamma_3$, which results in $\phi_e \approx \phi_s$, instead of $\phi_e \approx \Delta \phi +\phi_s$.
It means that it is no longer needed to set large $k_1$ to let $\hat \phi$ catch up with the dynamic $\phi$. 
Hence, $k_1$ can be much smaller that the PLL is more robust to the noises.
In Figure \ref{fig:pllc}, we choose the second-order PLL by setting $k_2\not = 0$.
Meanwhile, $k_1$ is much smaller than the one in Figure \ref{fig:plla} that the PLL is more robust to noises and does not cause observable jitters.
$k_1$ equals to the one in Figure $\ref{fig:pllb}$, but has no problem of catching up with the fast displacement for $k_2\not = 0$.
The accumulate displacement error is less than $2cm$ which is about at least 9 times more accurate than the one in Figure \ref{fig:plla}.



\section{Positioning by Synchronization}
\label{sec:sync}


Though we synthesize all the walking segments when user walks and turns, the problem is that the method has accumulated errors when we estimate the latter position by using the previous position, estimated walking directions and walking steps.
Especially when the user is far away and loses the signal from the speaker for a long time, the error increases and the historical measured position can no longer be used.
To solve this problem, we propose a synchronization mechanism that we leverage historical measurement to improve the robustness of \ourprotocol.



In synchronization, we additionally encode periodical pulses $s_2(t)$ in sending signal and propose the demodulating method to detect the receiving time of the pulses.
Since the pulses are periodical, the sending time of latter pulses can be predicted, if we can accurately estimate the sending time of one periodical pulse.
Hence, by using samples which can be directly used to calculate accurate position, we obtain the estimated distance, which infers traveling time $t_l$ from the speaker to the phone.
Then, we detect the receiving time $\tau'$ of one pulse in these samples and get the accurate sending time of the pulse $\tau=\tau'-t_l$.
Furthermore, the sending time of latter $i$th pulse equals to $\tau_i=\tau+iT$, where $T$ is denoted as period of pulses.
Hence, on obtaining the receiving time of $i$th pulse $\tau_i'$, we finally obtain real-time distance by using $\tau_i$ and $\tau_i'$ instead of the estimation method in Section \ref{sec:posest}.

\subsection{Pulse Modulation}
\label{sec:aaaa}

To design synchronization pulses $s_2(t)$ and the detection algorithm, several problems should be addressed:
\begin{itemize}
     \item Each speaker should not take much acoustic bandwidth in order to support more speakers in the room.
        Hence, $s_1(t)$ and $s_2(t)$ should be at the same frequency band, otherwise additional bandwidth for $s_2(t)$ is needed.
        Moreover, bandwidth of $s_2(t)$ needs to be narrow. 
        However, it is challenging that $s_2(t)$ should occupy more bandwidth if it can be successfully detected.
    \item $s_2(t)$ can also be used for displacement tracking by PLL.
        Otherwise, PLL will lose phase locks when processing $s_2(t)$.
\end{itemize}

Based on these requirements, we design $s_2(t)$:

\begin{equation}
s_2(t)=\begin{cases}
    \cos(2\pi ft+ \pi\sin\frac{\pi(t-\tau_i)}{T_p}) &  \tau_i \le t  \le \tau_i+ T_p \\
    \cos(2\pi ft) & \textrm{otherwise}
\end{cases}
\end{equation}
where we construct pulses starting at $\tau_1,  \dots, \tau_i$, and the duration of each pulse is $T_p$.


\begin{figure}[htpb]
    \begin{subfigure}[b]{0.5\textwidth}
        \begin{center}
            \includegraphics[width=2.5in]{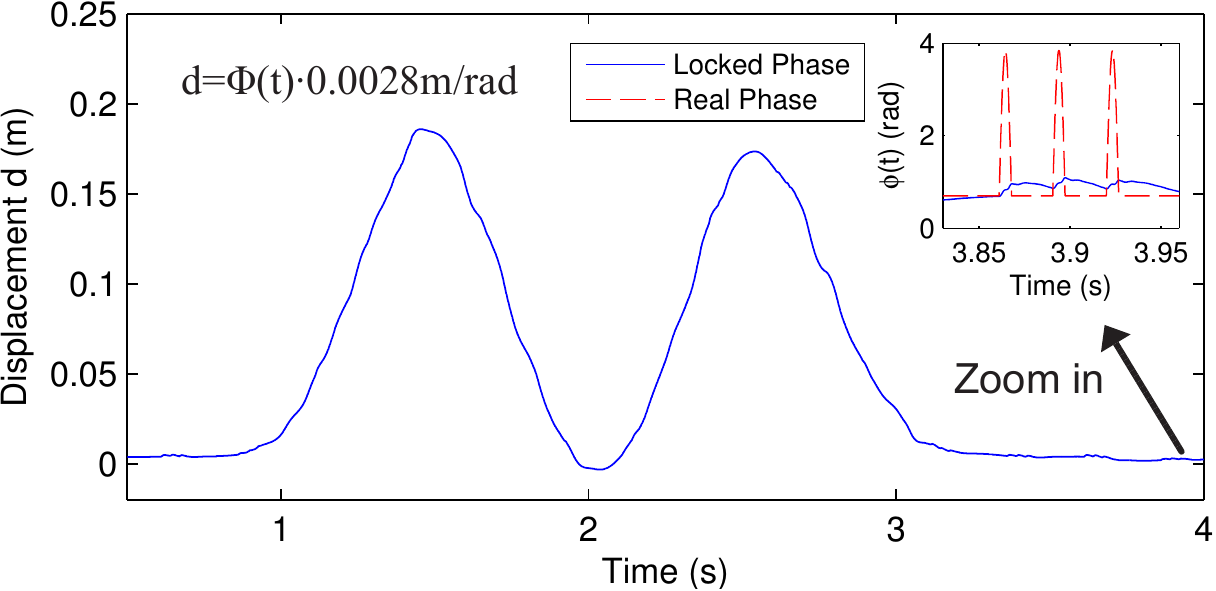}
        \end{center}
    \vspace{-0.1in}
        \caption{Measured Displacement}
        \label{fig:fthetagood}
    \end{subfigure}
    \begin{subfigure}[b]{0.5\textwidth}
        \begin{center}
            \includegraphics[width=2.5in]{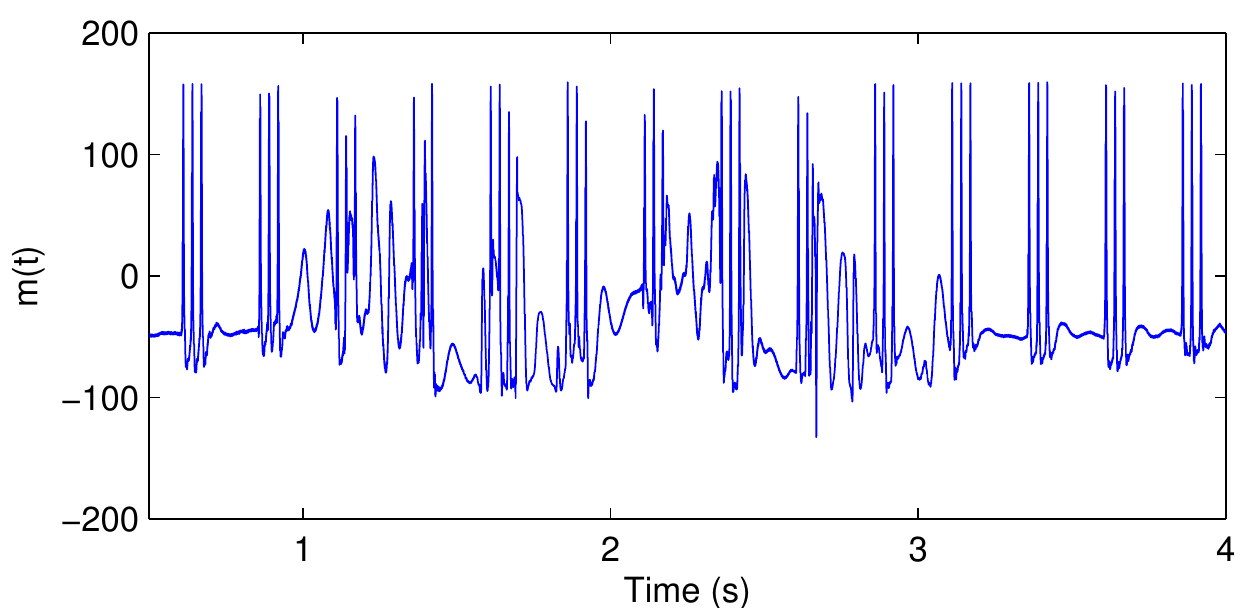}
        \end{center}
    \vspace{-0.1in}
        \caption{Detected Pulses}
        \label{fig:fmtgood}
    \end{subfigure}
    \caption{Calculated displacement and pulses from the same signal. }
    \label{fig:thetam}
    \vspace{-0.2in}
\end{figure}
More specifically, Figure \ref{fig:thetam} 
shows an example of detected pulse when the user moves the phone forward and backward twice and then stops.
We encode three adjacent pulses per $T_2=0.25s$.
Three adjacent pulses can be seen as a compensated periodical pulse with the period $T=T_2=0.25s$.
The time difference of the adjacent pulses is $T_3=0.03s$.
In Figure \ref{fig:fthetagood}, the estimated displacement is smooth and have no jitters whenever the phone is static or moving. 
We zoom in the calculated phase to show the performance of PLL when there are pulses in $s_2(t)$:
the calculated phase is not locked to the real phase;
instead, it seems that PLL has not detected the pulses that the phase is very smooth.
Specifically, while the maximum variation of the real phase is $\pi$, the corresponding variation computed by PLL is less than 0.4rad, which corresponds to the displacement of about $1mm$.
The cause of the phenomenon is that the parameters ($k_1$, $k_2$) of PLL are very small, and does track the fast changing phase.
Moreover, as the phase at the beginning of a pulse equals to the one at the end and the variation by PLL is small, the tracked phase finally becomes stable and the phase is the same with the one at the begining.

\subsubsection{Proof on Properties of Modulated Pulse}
We prove that the pulse $s_2$ does not take much acoustic bandwidth; and has little effects on the result of displacement tracking by PLL.

First, the central frequency of $s_2$  is the same as the one of $s_1$, except that the phase changes when there is a pulse.
Hence, $s_1$ and $s_2$ share the same frequency band.
Second, since the bandwidth of the pulse is about $\frac{\pi}{T_p}$ \cite{DBLP:journals/tim/SahuG08}, 
we set $T_p=0.007s$ so that the bandwidth is about 460Hz.
As the minimum frequency is 17000Hz when the acoustic is non-audible, and the maximum frequency which is supported by the phone is 24000Hz, 
the maximum concurrent signals that \ourprotocol supports in one place is $(24000-17000)/460 \approx 15$.
Actually, if the pulse has more narrow bandwidth, \ourprotocol will support more concurrent signals, whereas the pulse becomes harder to be detected.
How to modulate signals with more narrow bandwidth and demodulate the signal more accurately is left for future work.
Third, the component $s_3(t)=\pi\sin\frac{\pi(t-\tau_i)}{T_p}$ is the phase shift of the sine signal.
Furthermore, $s_3(t)$ starts and ends at the same value $0$, and the maximum value of $s_3$ is $\pi$.
Hence, the displacement will not be affected by the pulse theoretically.

\subsubsection{Discussions of Pulse Modulation}
\textbf{Choosing Parameters:}
There is a trade off on choosing the parameters $T_p$, $T_1$, $T_2$, $T_3$, we show the analysis on choosing the parameters as follows:
\begin{inparaenum}[\itshape a\upshape)]
    \\\indent\item{$T_p$:} As the bandwidth of pulses equals to $\frac{\pi}{T_p}$, smaller $T_p$ results in wider bandwidth requirement and less simulateneous signals in the same room.
        On the other hand, greater $T_p$ results in less accuracy of displacement tracking. 
        The reason is that the pulses are regarded as noises in displacement tracking.
    \\\indent\item{$T_1$:} Since there are 3 adjcented pulses in one compensated periodical pulse, $T_1=T_2-3T_3$. 
    \\\indent\item{$T_2$:} Recall that $T_2=T$ which is the period of compensated pulses. 
        Smaller $T_2$ will enhance the accuracy of measuring the receiving time of pulses for we have more pulses for matching.
        However, if we choose smaller $T_2$, we may face the ambiguity problem. 
        Specifically, denote the receiving time of a pulse is $t_r$ and the sending time of periodical pulses is $t_s+kT_2$.
        The calculated distance is $v_a(t_r-t_s-kT_2)$, where $k$ is an undetermined integer which also makes the distance undetermined.
        To get $k$, we further leverage the maxmium distance from speaker to anchor, denoted as $l_m$.
        Since $v_a(t_r-t_s-kT_2)<l_m$, to get the unique solution of distance, $v_aT_2$ should be greater than $l_m$. 
        In our paper, we assume that $l_m=85m$ which infers $T_2=0.25s$.
    \\\indent\item{$T_3$:} $T_3$ has limitation on its minimum value. Firstly, to avoid overlaps of adjacent pulses, $T_3>T_p$. 
        Secondly, there also should be intervals between adjacent pulses. 
        We zoom in Figure \ref{fig:fthetagood} and find that PLL needs time longer than the duration of pulses to lock the displacement to the real value after the pulse terminates.
        Hence, if adjcent pulses are too close, PLL may become very unstable.
        If $T_3$ increases, for $T_1=T_2-3T_3>0$, $T_2$ may also increase which also affects the performance of \ourprotocol.
\end{inparaenum}

\noindent\textbf{Reducing Signal Conflicts:}
As explained earlier, due to the bandwidth limitation, our default parameter of pulse modulation supports 15 concurrent signals. 
Here, to reduce signal conflicts, we find that further optimizations can be made for different applications as follows:
\begin{inparaenum}[\itshape a\upshape)]
    \\\indent\item 
Virtual business card sharing: In this case, users are usually close to each other, and we can choose to narrow the bandwidth of pulses in synchronization.
Hence, \ourprotocol can support more users who broadcast signals simultaneously, while we only reduce the accuracy of pulse detection and long distance positioning, which are not much required.
\\\indent\item
Virtual shopping guide:
We suggest that if there is requirement of more shopping guides, we can use only a few speakers for normal indoor localization, instead of just relative positioning.
Our further evaluations in Section \ref{subsec:all} prove that \ourprotocol supports unlimited number of shopping guides by simple and sparse deployment of speakers, \ie, the smart device only receives signals from 2 speakers on average, but gains 1-meter accuracy.
\end{inparaenum}

\subsection{Pulse Detection}
We discuss how we detect the receiving time $\tau_i'=\tau_i+t_l$ of the $i$th pulse by leveraging the component $s_3(t)$.
Assuming the locked phase by PLL is $\phi_r$ before the pulse starts, the expected pulse is $\tilde r(t)=\cos(2\pi f t+\phi_r+\pi \sin \frac{\pi(t-\tau_i')}{T_p})$.
Hence, for the received sample $r(kT_s)$, we compute the likelihood $m(kT_s)=\sum_{i=k}^{k+T_p/T_s}r(iT_s)\tilde r (iT_s) $, \ie, when $m(kT_s)$ reaches the maximum, the corresponding $kT_s$ is the starting time of the received pulse.
Note that, if we set expected pulse $\hat r(t)=\cos(2\pi f t+\phi_r+\pi)$ and there is no pulse for the next $T_p$ that $r(t)=\cos(2\pi f t+\phi_r)$, $\hat m(kT_s)=\sum_{i=k}^{k+T_p/T_s}r(iT_s) \hat r (iT_s) $ will reach the minimum. 
Actually, $s_3(t)$ is the filtered version of pulse $\hat s_3(t)=\pi$ that the pulse $s_3(t)$ has narrower bandwidth.
Accordingly, $\tilde r(t) \approx \hat r(t)$ which means $m(t)$ will reach the value close to minimum when there is no pulse in the next $T_p$. 
Hence, arrival time $\tau_i'$ of the shape can be detected by $m(t)$.



\subsubsection{Analysis on Design of Pulse Detection}
As mentioned earlier, our PLL takes $s_2(t)$ as noises and only tracks $s_1(t)$.
There are two advantages based on above results: 
1) the pulses have very small effects on the tracked displacement. 
2) For the variation is very small and the variation of $\phi_r$ is stable when there are pulses, peaks of $m(t)$ become clear to be detected. 
In Figure \ref{fig:fmtgood},  $m(t)$ reaches the peak value (\ie, 150),  when there is a pulse at $t$ and the bottom value (\ie, -50) when there are almost no pulses.
As a whole, it shows an interesting result that on demodulating $s(t)$, the peak of $m(t)$ is very clear for synchronization in Figure \ref{fig:fmtgood}, while the corresponding calculated phase is very smooth for displacement tracking in Figure \ref{fig:fthetagood}.

We can also find that when the phone is static, the peaks corresponding to the pulses are clear.
However, they are unclear when the phone is moving.
Furthermore, when the signal is weak, the periodical peaks cannot be detected by $m(t)$ in Figure \ref{fig:fmtbad} due to noises.
Hence, we make further solution to make the peaks more clear in case that the phone moves or the signal is weak.
The solution is based on the observation that expected peaks still appear at expected time, though they sink in the noises.
Meanwhile, random peaks have fewer chances to appear periodically.
Hence, we assign $m_1(t)=m(t-T_3)+m(t)+m(t+T_3)$ in Figure \ref{fig:fmtbad1}, where the peaks are more clear to be identified in $m_1(t)$.
Then, we assign $m_2(t)=m_1(t-T_2)+m_1(t)+m_1(t+T_2)$ in Figure \ref{fig:fmtbad2}, where the peaks can be easily detected.
Moreover,  when the phone is moving and the corresponding phase is in Figure \ref{fig:fthetagood}, the peaks are also very clear in Figure \ref{fig:fmtgood2}.

\subsubsection{Dealing With Multipath Effects:}
We also find that the result of synchronization is affected by multipath effects, especially when the smart device is static.
Hence, we make further study and improvement on pulse detection.

\begin{figure}[htpb]

   \vspace{-0.1in}
   \begin{subfigure}[b]{0.23\textwidth}
       \begin{center}
           \includegraphics[width=1.7in]{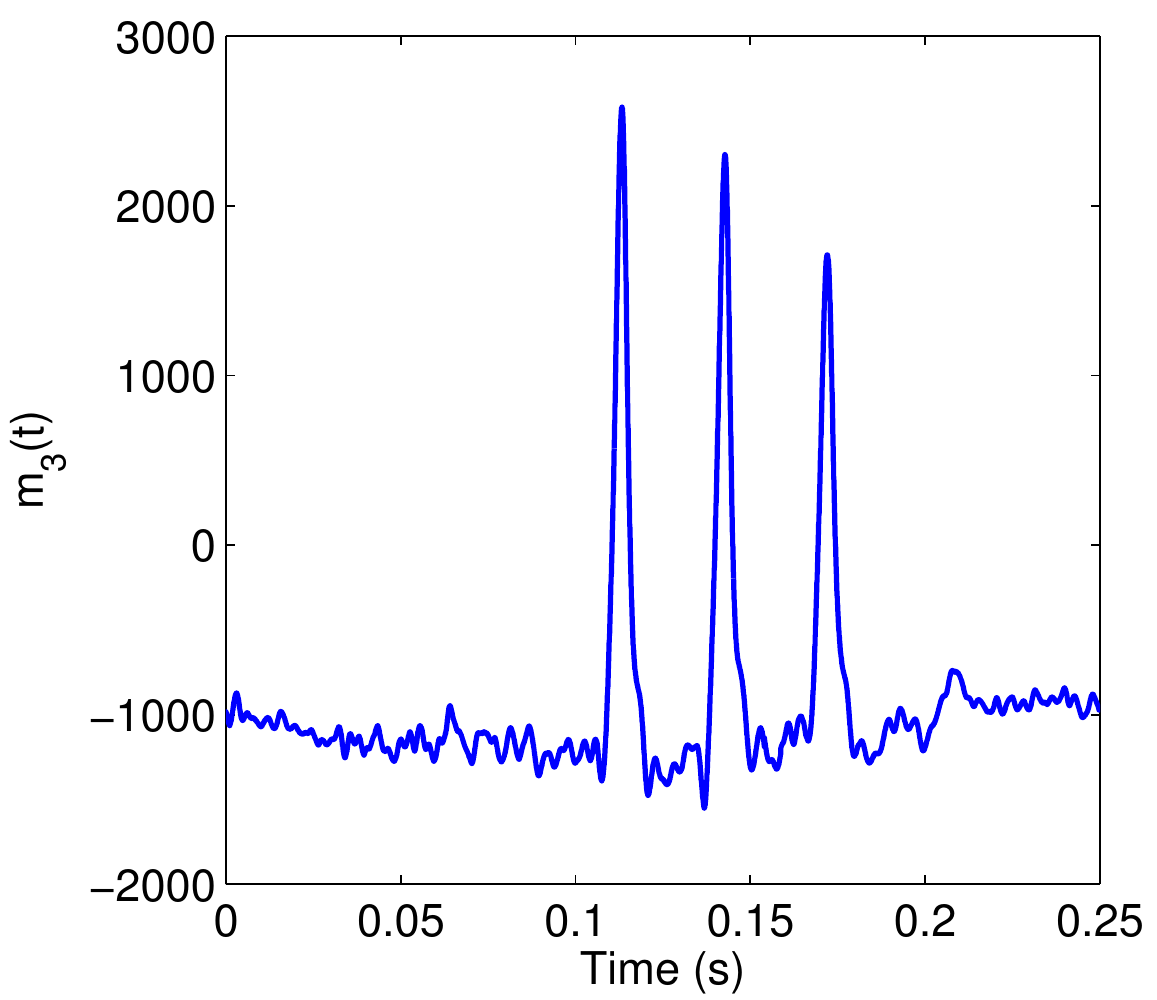}
       \end{center}
       \caption{Good Case.}
       \label{fig:multipathgood}
   \end{subfigure}
   \begin{subfigure}[b]{0.23\textwidth}
       \begin{center}
           \includegraphics[width=1.7in]{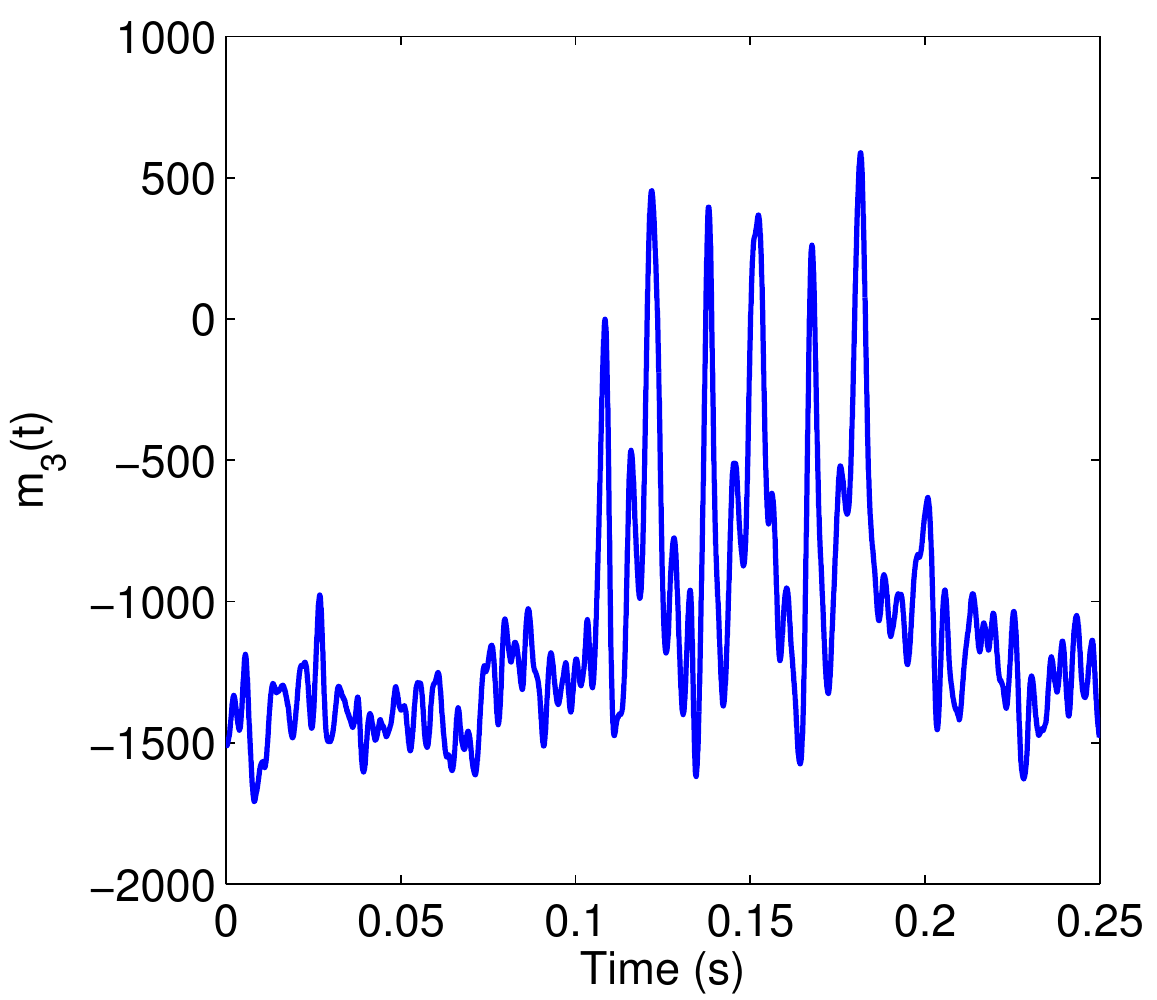}
       \end{center}
       \caption{Bad Case (Multipath).}
       \label{fig:multipathbad}
   \end{subfigure}
   \caption{Pulse detection in case of multipath effects.}
   \label{fig:multipath}
   \vspace{-0.2in}
\end{figure}

We find that when the phone is static, there is another property which can be leveraged: the distance from the smart device to the dummy speaker is constant. 
Hence, we can use $m_3(kT_2)=\sum_{i\in\{x|x=k \mod T_2\}}m(iT_2)$, which sums all the $m_3(t)$ of pulses and make the detected time of pulses more clear.
The result of summed $m_3(kT)$ is shown in Figure \ref{fig:multipath}.
In Figure \ref{fig:multipathgood} when there is no multipath effect, there are 3 pulses in a period $T_2$. 
However, in Figure \ref{fig:multipathbad}, which is gathered from the shopping mall, there are 9 pulses at least, which means there are 2 additional paths reflected from walls or other objects.
In this case, all the 3 paths are the possible pulses directly received from the dummy speaker.

After recognizing the possible multipaths, we make further step to filter the direct path.
Specifically, we use the result of PLL, which corresponds to the displacement.
As displacement tracking is less affected by multipath effects, we compare with the result of PLL and pulse detection when a user walks from one position and stops at another one.
In this case, denote that the displacement by PLL is $d$, and the receiving time of pulses are in the set 
$T_a=\{t_{a1},t_{a2},t_{a3},\dots\}$ and
$T_b=\{t_{b1}, t_{b2},t_{b2},\dots\}$ at the start point and end point respectively. 
Hence we obtain the receiving time $(t_a, t_b)=\underset{t_a \in T_a, t_b \in T_b}{\arg \min} |(t_a-t_b)v_a-d|$.

\subsection{Positioning after Synchronization}
Assume sending time of the next pulse is $t_s$, which is the result of synthesizing and the detected receiving time is $t_r$. 
Then distance $l=v_a(t_r-t_s)$ and the distance at the horizontal plane is $L=\sqrt{l^2-h^2}$.
For direction estimation, we first calculate $x$ and $y$ using newly obtained $L$, previous $s$ and $d_i$.
For example, on calculating $\psi_1'$ in Figure \ref{fig:distance}, assume $l_1$ is obtained from synthesizing. 
Since $x=-l_1\cos\psi_1$ and $y=l_1\sin\psi_1$, $l_i'$ in \eqqref{eq:mine0} has the following form
\begin{equation}
    l_i'=\sqrt{l_1^2 \sin^2 \psi_1+(-l_1\cos \psi_1 +(i-1)s)^2}
    \label{eq:mine1}
\end{equation}
Hence, $\psi_1$ is obtained by $ \underset{\psi_1}{\arg \min} \sum_{i=1}^ne_i^2$ where $e_i$ is calculated by \eqqref{eq:mine}, \eqref{eq:mine1}.
Then $\cos\psi_1'= \frac{l_1 \cos \psi_1}{\sqrt{l_1^2-h^2}}$.

\section{Performance Evaluation}
\label{sec:exp}
\begin{figure*}[t]
    \centering
    \begin{subfigure}[b]{0.16\textwidth}
        \begin{center}
            \includegraphics[width=1.2in]{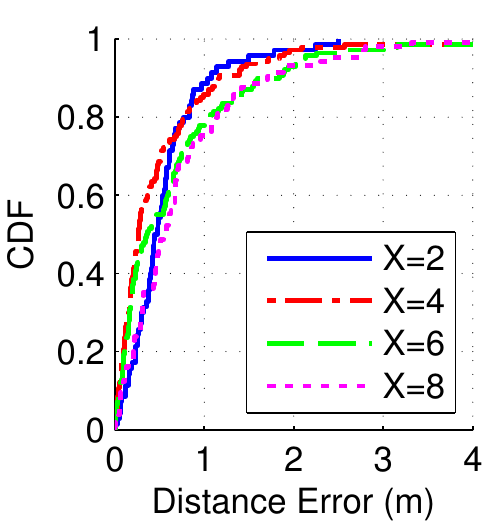}
        \end{center}
    \vspace{-0.1in}
        \caption{ Ranging }
        \label{fig:perrorlx}
    \end{subfigure}
    \begin{subfigure}[b]{0.16\textwidth}
        \begin{center}
            \includegraphics[width=1.2in]{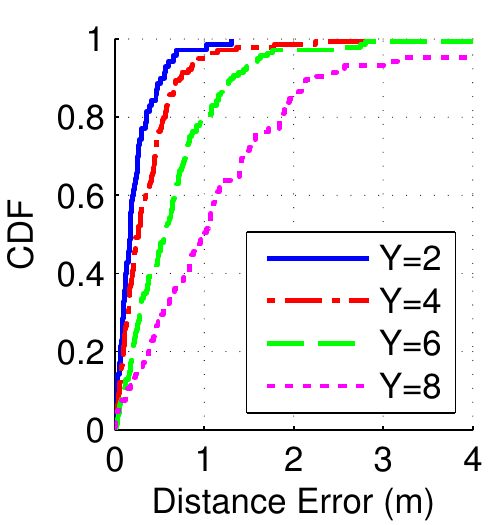}
        \end{center}
    \vspace{-0.1in}
        \caption{Ranging }
        \label{fig:perrorly}
    \end{subfigure}
    \begin{subfigure}[b]{0.16\textwidth}
        \begin{center}
            \includegraphics[width=1.2in]{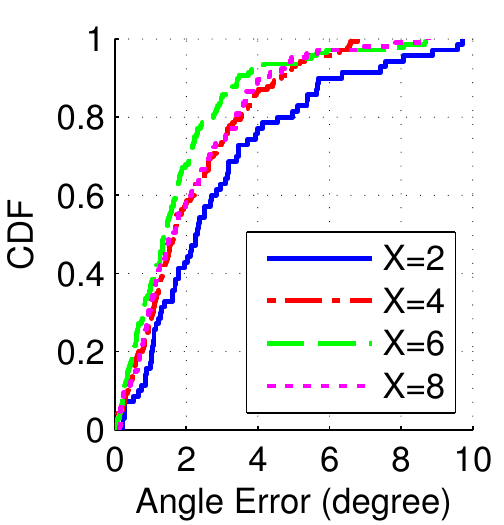}
        \end{center}
    \vspace{-0.1in}
        \caption{Direction}
        \label{fig:perrorax}
    \end{subfigure}
    \begin{subfigure}[b]{0.16\textwidth}
        \begin{center}
            \includegraphics[width=1.2in]{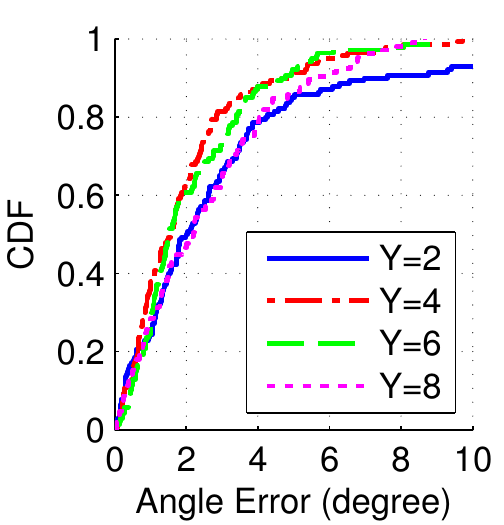}
        \end{center}
    \vspace{-0.1in}
        \caption{Direction}
        \label{fig:perroray}
    \end{subfigure}
    \begin{subfigure}[b]{0.16\textwidth}
        \begin{center}
            \includegraphics[width=1.2in]{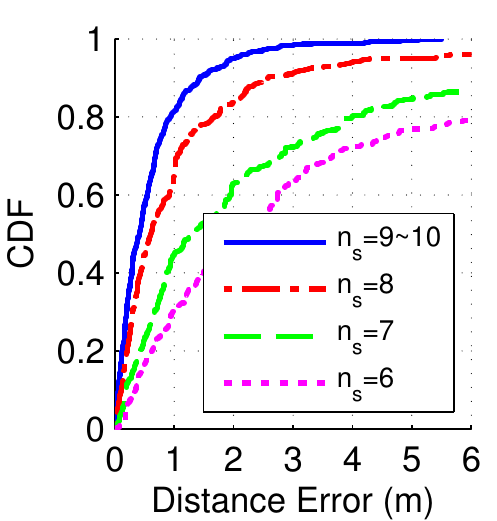}
        \end{center}
    \vspace{-0.1in}
        \caption{ Ranging }
        \label{fig:steplerror}
    \end{subfigure}
    \begin{subfigure}[b]{0.16\textwidth}
        \begin{center}
            \includegraphics[width=1.2in]{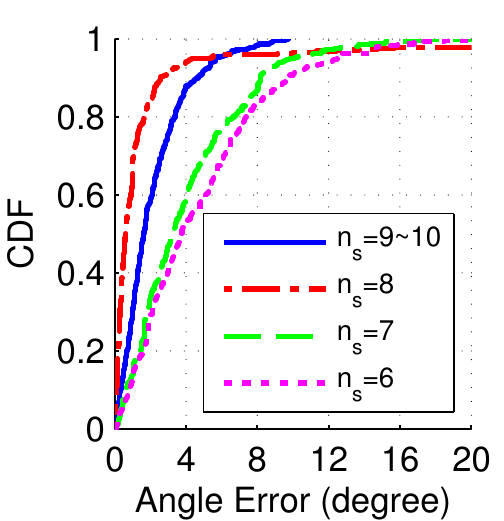}
        \end{center}
    \vspace{-0.1in}
        \caption{Direction}
        \label{fig:stepaerror}
    \end{subfigure}
    \label{fig:steperror}

\caption{The accuracy of ranging and direction finding 1) when the user starts walking at different positions (a)(b)(c)(d), 2) when the user walks for smaller number of steps (e)(f). }
    \vspace{-0.2in}
\end{figure*}



In this section, we perform system evaluation by using two types of speakers: Samsung Galaxy Note 2 and normal dummy speakers.
The speaker merely broadcasts acoustic waves and does not perform communications.
We mainly use Google Nexus 4 to receive the acoustic signals.
We do not make any modifications to the phone or jailbreak the operation system, and all the components, such as BPF, AGC, PLL, are implemented by the software.
We evaluate the performance in an empty room, an office and the shopping mall.
The micro benchmarks are made for position estimation and synchronization. 
We then evaluate the total performance where all the components are used.

Note that in our system, we do not measure the walking direction in World Coordinate System (WCS).
The main reason is that this measurement is not necessarily needed in our system and we only calculate the relative position between walkers and speaker.
Furthermore, accurately measuring walking direction in WCS is still challenging \cite{2012-MOBICOM-Zeezeroeffort}, which is mainly caused by unpredictable errors when using compass, especially in indoor environment.
Therefore, to evaluate the accuracy of our system, we build and rely on relative coordinate system (RCS), instead of classical World Coordinate System (WCS).
In RCS, we set the direction of the piecewise linear segment as X axis, and starting point of the segment as origin point.
For instance, assuming $Y=\sqrt{y^2-h^2}$ and $X=-x$ in Figure \ref{fig:distance}, $(X,Y)$ is the position of the speaker when the user starts walking.


\begin{figure*}[htpb]
    \centering
    \begin{subfigure}[b]{0.155\textwidth}
        \begin{center}
            \includegraphics[height=1.1in]{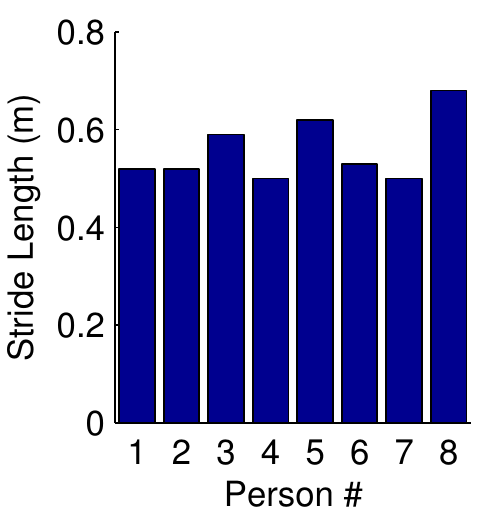}
        \end{center}
        \vspace{-0.1in}
        \caption{ Stride Length}
        \label{fig:personstep}
    \end{subfigure}
    \begin{subfigure}[b]{0.145\textwidth}
        \begin{center}
            \includegraphics[height=1.05in]{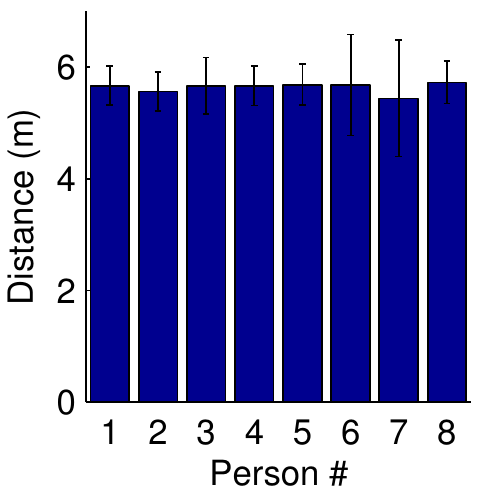}
        \end{center}
        \vspace{-0.1in}
        \caption{ Ranging }
        \label{fig:personL}
    \end{subfigure}
    \begin{subfigure}[b]{0.145\textwidth}
        \begin{center}
            \includegraphics[height=1.1in]{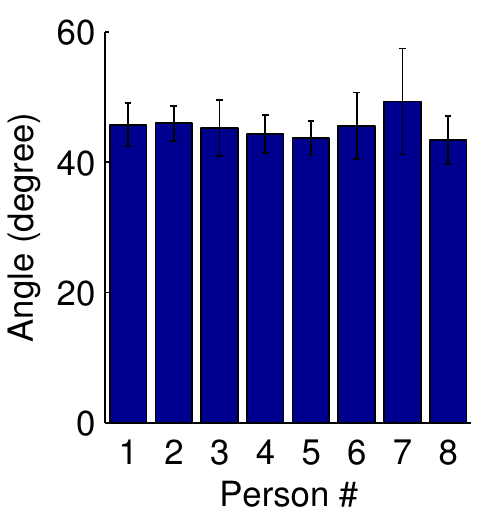}
        \end{center}
        \vspace{-0.1in}
        \caption{Direction }
        \label{fig:personA}
    \end{subfigure}
    \begin{subfigure}[b]{0.12\textwidth}
        \begin{center}
            \includegraphics[height=1.1in]{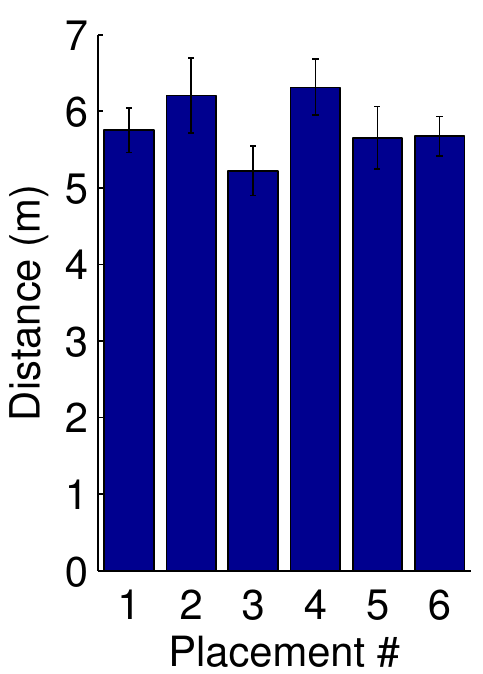}
        \end{center}
        \vspace{-0.1in}
        \caption{ Ranging }
        \label{fig:placementL}
    \end{subfigure}
    \begin{subfigure}[b]{0.12\textwidth}
        \begin{center}
            \includegraphics[height=1.1in]{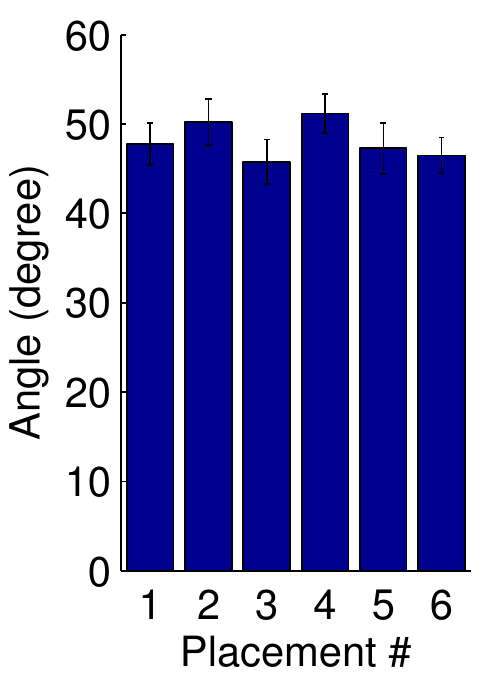}
        \end{center}
        \vspace{-0.1in}
        \caption{Direction }
        \label{fig:placementA}
    \end{subfigure}
    \begin{subfigure}[b]{0.10\textwidth}
        \begin{center}
            \includegraphics[height=1.1in]{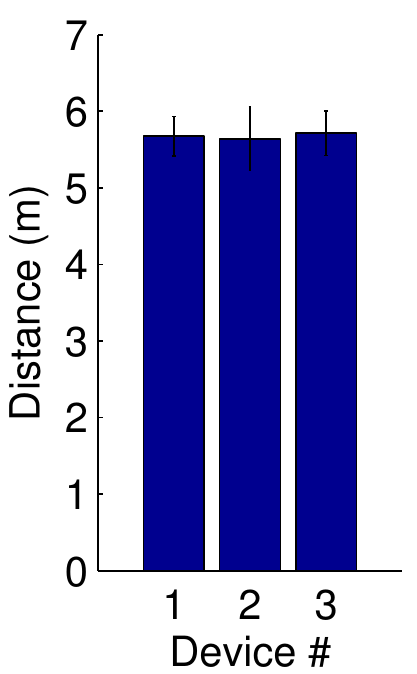}
        \end{center}
        \vspace{-0.1in}
        \caption{ Ranging }
        \label{fig:deviceL}
    \end{subfigure}
    \begin{subfigure}[b]{0.10\textwidth}
        \begin{center}
            \includegraphics[height=1.1in]{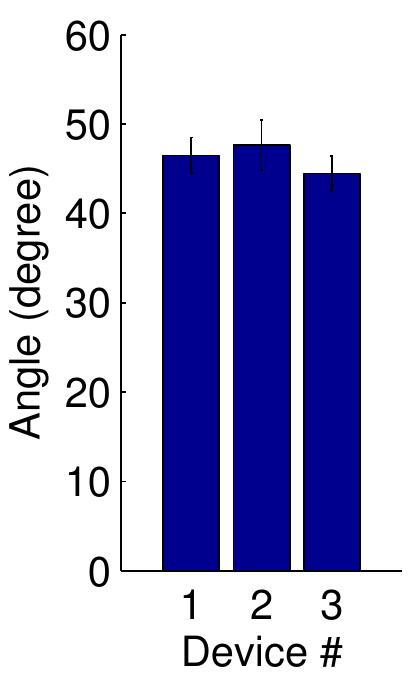}
        \end{center}
        \vspace{-0.1in}
        \caption{Direction }
        \label{fig:deviceA}
    \end{subfigure}
    \vspace{-0.1in}
    \caption{The mean and standard deviation of ranging and direction estimation for different users (a)(b)(c), placements (d)(e), and smart devices (f)(g). }
    \label{fig:expperson}
    \vspace{-0.2in}
\end{figure*}

\subsection{Position Estimation}
We evaluate position estimation in several types of cases, \ie, different related positions from the phone to the speaker, number of walking steps, users, orientation of devices, device diversity and environments, which may affect accuracy of the estimation.
\subsubsection{Positions}
\label{sec:casepos}

We make evaluation in an empty room to evaluate the performance at different places.
In this experiment, the speaker is placed at different locations, \ie, $X=2,4,6,8m$, and $Y=2,4,6,8m$.
We let the user walk for $9\sim 10$ steps with the walking lengths of about $6m$.
The relative height $h$ is about $0.3m$.
For each location, the user holds the phone in hand and walks for 35 times to gather samples, \ie, we get 560 samples in this micro benchmark.
Note that by using other smart devices, such as smart glasses, or smart watches, the user can have more comfortable experience.
Due to that we only use IMU sensors and microphone which are frequently used in smart devices, we use the smartphone as smart device in the experiment.
Then, we calculate the relative position for evaluating the accuracy of calculated distance $L=\sqrt{X^2+Y^2}$ and direction $\psi'=\arccos(X/L)$.
Note that since the user walks for only several steps and the walking distance is short, we only evaluate the accuracy of the initial position $(X,Y)$.

In Figure \ref{fig:perrorlx}, the accuracy of distance estimation is very close for different $X$.
We further study the distribution of large errors in Figure \ref{fig:perrorly}.
We find an interesting fact that the errors are nearly proportional to $Y$.
Hence, when $Y=2,4,6,8m$, the corresponding errors are within $0.35m$, $0.55m$, $0.97m$, $1.88m$ at the percentage of $80\%$.
The result is acceptable in our case for the user requires higher level of accuracy when s/he is close to the speaker.
Furthermore, we use synchronization and synthesizing scheme achieves the accurate ranging in longer distances, instead of this position estimation scheme.


For direction estimation, it is still very accurate when the $X$ or $Y$ increases in Figure \ref{fig:perrorax}, \ref{fig:perroray}.
As a total, the mean of ranging and angle error is $0.63m$ and $2.46^o$ respectively. 

\subsubsection{Number of Steps}
The accuracy of the position estimation depends on number of walking steps.
We compare the results when the user walks for smaller number of steps $n_s$ in Figure \ref{fig:steplerror}, \ref{fig:stepaerror}.
The samples are the same ones gathered in section \ref{sec:casepos}, and the only difference is that we only use part of each sample which infers fewer walking steps.
The results show that the ranging errors increase quickly when $n_s$ reduces.
The reasons are:
1) The user's stride length varies occasionally. 
2) User's phone also shifts left and right regularly, \ie, it does not move strictly in a line, when the user holds the phone and walks. 
As these facts will have less effects on the accuracy when $n_s$ is larger, it can be foreseen that the accuracy will continue to be improved  when $n_s>10$, though it is already very accurate when $n_s=10$.


    



The estimated direction is also affected by the smaller $n_s$ in Figure \ref{fig:stepaerror}.
But it is still acceptable that the angle errors are under $8^o$ at the percentage of $80\%$, when $n_s=6$.
As a whole, when $n_s$ is small, the accuracy is enough for direction estimation of surrounding speaker, while it requires latter synthesizing scheme to obtain accurate distance.

\subsubsection{Users}
Different user has different stride length and user motion when the user walks, which causes variation of displacement patterns displacement $d_i$ and might affect the positioning result.
Hence, we recruit 8 volunteers in this experiment: each user walks in a line of about $6m$ for 35 times where $(X,Y)=(4,4)$.

We have the following observations in Figure \ref{fig:expperson}: The standard deviations (std) of the ranging and direction are small for most users.
In Figure \ref{fig:personstep}, the person 1,2,4,6,7 have small stride lengths while the rest ones have bigger length,
but the result is similar for all the users (except for the person 6,7).

The results infer that the stride length is very stable and the positioning accuracy is not much affected by variation of stride length, though the stride length between different users may be much different.

\subsubsection{Orientation of Speaker and Microphone}
We consider the cases when the speaker or the microphone faces to different directions: 
(1) (default) the microphone faces to the sky, and the speaker faces to the walking line.
(2) microphone, facing to the front.
(3) microphone, perpendicular to the walking direction and facing to the speaker.
(4) microphone, facing to the ground.
(5) microphone, perpendicular to the walking direction and speaker is at the back of the microphone.
(6) speaker, facing to the ground.
The result in Figure \ref{fig:placementL}, \ref{fig:placementA} shows that the std is small in all cases and the result is very stable. 

We also find that the mean value of distance increases when the signal is weaker in case (2), (4) and decreases when signal is stronger in case (3).
The reason is that when the signal is weak, PLL will lose some signals and the displacement decreases, which makes the calculated distance become larger. 
Hence, based on our measurements in displacement tracking, we make calibrations on the calculated PLL. 
More specifically in case (1) that the displacement $d= 1.22 \frac{v_a}{2\pi f}\Delta\phi$; if $d>0$, and $d= 1.69 \frac{v_a}{2\pi f}\Delta\phi$, if $d<0$, where $\Delta \phi$ is the tracked phase shift.
Note that we make calibration with constant factor (\ie, 1.22), for the environment has limited effect on the result of PLL when the signal is strong enough.
However, when $d<0$, which means the speaker is at the back of the walking user, $d$ is usually not used for position estimation if the tracked phase is abnormal (\eg, when \ourprotocol cannot detect pulses from the phase).

\subsubsection{Device Diversity}
We test several Commercial Off-the-Shelf (COTS) smart devices as acoustic receivers: (1) Nexus 4, (2) Samsung Galaxy Note 2. (3) Nexus 7. 
We choose $(X,Y)=(4,4)$ as the start point of walking, and the error of position estimation is shown in Figure \ref{fig:deviceL}, \ref{fig:deviceA}. 
The result shows that these smart devices have similar performance.

We also use normal dummy speakers as acoustic speakers when we make experiment in a large shopping mall, for we consider the case that the normal speakers serve as virtual shopping guides.

\noindent \textbf{Calibration of clock drift:} We find some interesting phenomenon: different from the previous smart devices, the normal speaker has serious clock drift and needs to be calibrated. 
For instance, when a speaker is supposed to broadcast signal at 19000Hz, the actual received signal is 19007Hz.
If the frequency drift is 0.1Hz, the error of distance measuring is about 600*340*0.1/19000=1.07m, when the smart device performs synchronization for 10 minutes. 
To solve this problem, our design of PLL measures the precise clock offset when the receiver is static for only a few seconds.
In this case, $\gamma_2$ in Figure \ref{fig:pll} rapidly converges to a constant value.
As $\gamma_2$ equals to the phase shift per sampling time $T_s$, the frequency offset equals to $\frac{k_2}{2\pi T_s}$.
Hence, once we let the smart device be static for a few seconds, the precise frequency offset is obtained. Afterward, we calibrate the clock drift in real-time using the constant frequency offset and there is not any clock drift after calibration.

\subsubsection{Environments}
We compare the accuracy of position estimation in the empty room and at different locations in the office.
We find that it shows the similar results.
We further evaluate the effects in a shopping mall in the latter subsection.


\subsection{Synchronization}
\label{sec:synchronization}
\begin{figure}[htpb]
    \vspace{-0.15in}
    \begin{subfigure}[b]{0.235\textwidth}
        \begin{center}
            \includegraphics[width=1.7in]{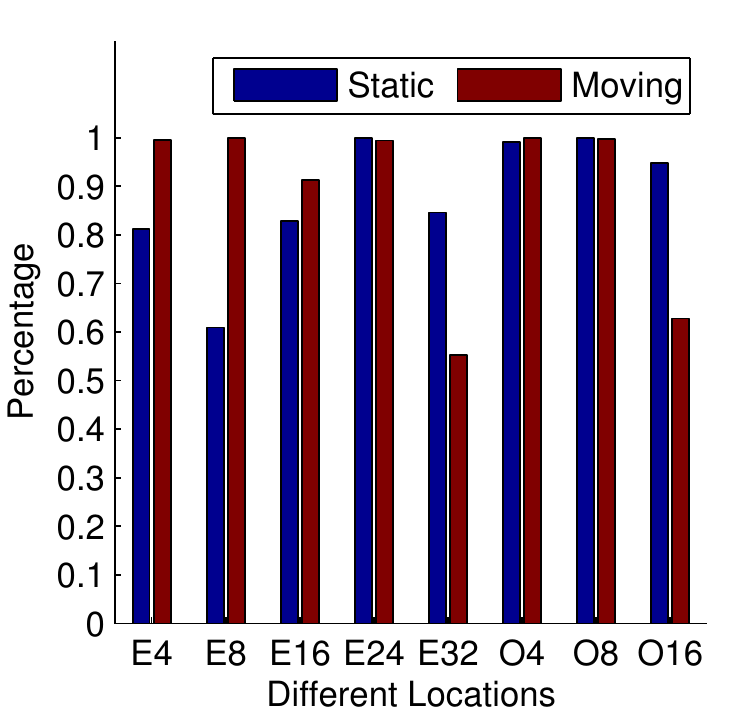}
        \end{center}
    \vspace{-0.1in}
        \caption{ $80cm$ criterion. }
        \label{fig:syncexp1}
    \end{subfigure}
    \begin{subfigure}[b]{0.235\textwidth}
        \begin{center}
            \includegraphics[width=1.7in]{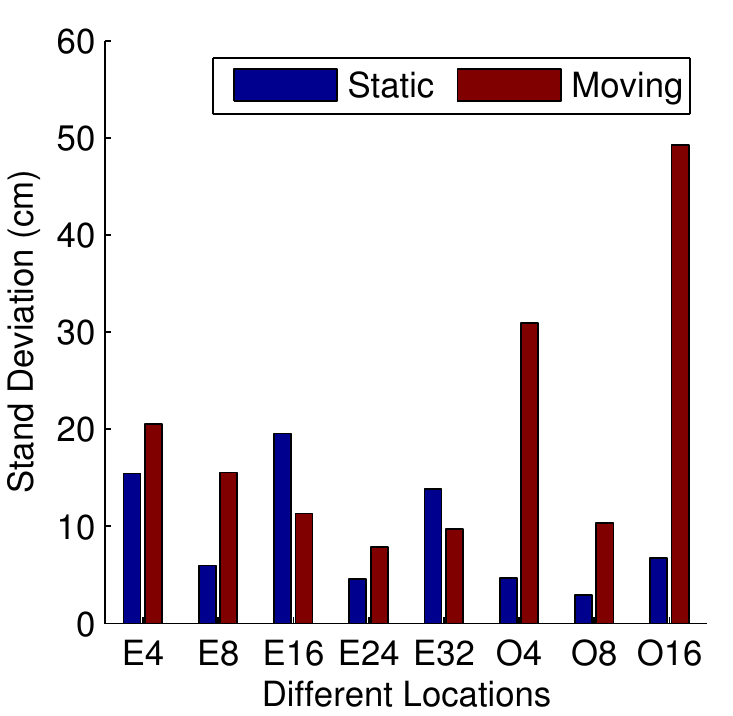}
        \end{center}
    \vspace{-0.1in}
        \caption{Stand Deviation}
        \label{fig:syncexp2}
    \end{subfigure}
    \caption{(a) Percentage of successful experiments at different locations (b) standard deviation.}
    \label{fig:syncexp}
    \vspace{-0.2in}
\end{figure}

In Figure \ref{fig:syncexp}, we choose 8 locations in an empty room and the office to evaluate the performance of synchronization.
For example, E32 means that the experiment is in the room and the distance from the smart device to the speaker is $32m$,
and O16 means that it is in the office and the distance is $16m$.
In each position we test two cases: the phone is static or moving back and forth without stop.
For each case, the phone records the audio for 100 seconds, which means there are 400 signals for synchronization in the samples. 
Then, we evaluate the accuracy of pulse detection.
For easier understanding of our results, the error of arrival time $t_e$ is converted to distance measurement error $l_e=v_at_e$.
For instance, if the error is the time interval of 1 acoustic sample, \ie, $t_e=\frac{1}{44100}s$, the corresponding distance error is $l_e\approx0.8cm$.

Since we find that there are occasional significant errors ($>3m$), we first set threshold $l_t=80cm$ and evaluate ratio of successful detection that $l_e<l_t$.
In Figure \ref{fig:syncexp1}, the successful detection rate is above $80\%$ for most cases when the phone is static.
When the phone is moving, the performance is good as well if the distance is within $24m$ and $8m$ in the empty room and office respectively.
In some cases the rate is close to $100\%$.

There is also an exception that at location E8 when the phone is static, the rate is only $61.0\%$, while it reaches $100\%$ at the same place when the phone is moving.
So, we conduct the experiment again at the same place, and the result is close to the previous one.
We suppose it is caused by the multipath effects: the phase $\phi_r$ changes according to the mixed signals and becomes stable when it is static, which affects the result of pulse matching.
The reason of high successful rate in case of moving phone is that: though it is also affected by multipath, the phases of reflected signal at different positions are irregular. 
In other words, the PLL locks the phase of the signal directly from the speaker, \ie, the multipath signal is regarded as noises by PLL.
Hence, the performance is better when the phone is moving.
We find the location E4, E8, E16 also have the same phenomenon, which validates our hypothesis.
Actually, this is a good result for \ourprotocol: when the user is walking, the synchronization result is very good and can be directly used for synthesizing; when the user is walking, as the successful detection rate is above $60 \%$, \ourprotocol collects enough samples and then determines the most possible receiving time.
In Figure \ref{fig:syncexp2}, we show the standard deviation of results in case of successful detection.
The std in most cases are around $10cm$ expect that the std is $30.9cm$ and $49.2cm$ when the phone is moving at O4 and O16 respectively. 


\subsection{Positioning after Synchronization}
\label{sec:putit}
We evaluate the performance of \ourprotocol which uses position estimation and synchronization in the following steps:
\begin{enumerate}
    \item The user walks in a line where the initial coordinate of the speaker is $(4,4)$.
        In this step, we calculate the distance through position estimation and then calculate the sending time of periodical signals $s_2(t)$ by synchronization.
\item The user then turns, walks and stops at the position where relative coordinate of speaker is $(X,Y)$.
\item The user walks again for about $6m$.
    The position, which is supposed to be $(X,Y)$, is then computed according to the sending time and the received samples in this short duration of walking.
\end{enumerate}
\begin{figure}[htpb]
    \vspace{-0.2in}
    \begin{subfigure}[b]{0.235\textwidth}
        \begin{center}
            \includegraphics[width=1.8in]{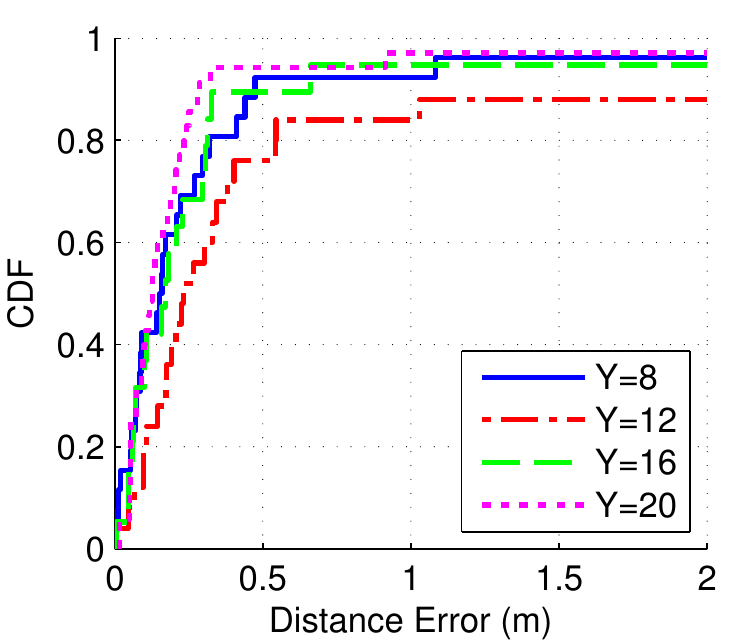}
        \end{center}
    \vspace{-0.1in}
        \caption{ Ranging }
        \label{fig:totallerror}
    \end{subfigure}
    \begin{subfigure}[b]{0.235\textwidth}
        \begin{center}
            \includegraphics[width=1.8in]{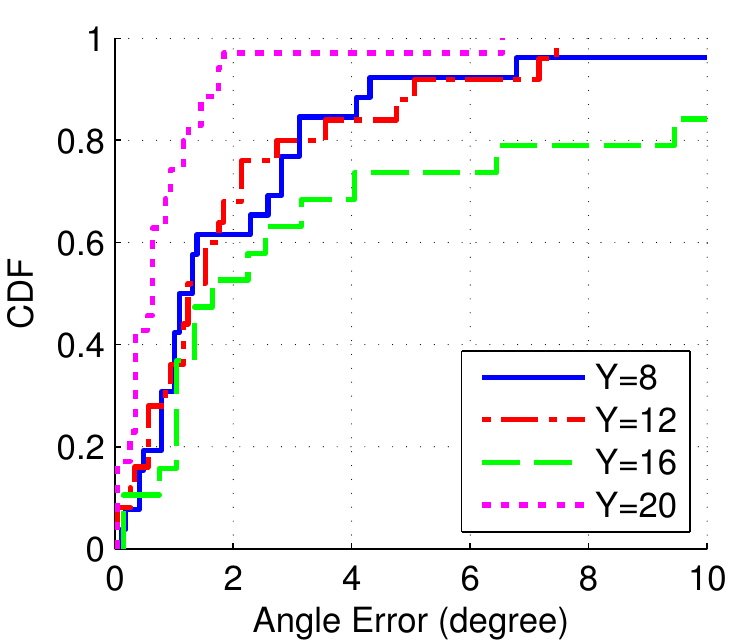}
        \end{center}
    \vspace{-0.1in}
        \caption{Direction Estimation}
        \label{fig:totalaerror}
    \end{subfigure}
    \caption{Accuracy of positioning by synchronization. }
    \vspace{-0.15in}
    \label{fig:totalerror}
\end{figure}

We conduct the experiment in the empty room and the office.
Specifically, we set $(X,Y)=(4,12)$ and $(4,20)$ in the empty room to gather the samples and $(4,8)$ and $(4,16)$ in the office.

In Figure \ref{fig:totallerror}, the ranging errors are under $0.32m$ and $0.66m$ at the percentage of $80\%$ and $90\%$ for most cases.
It means that both position estimation and synchronization achieve considerable accuracy.
There are also occasional errors for each cases which are greater than $2m$.
It is caused by the multipath effects in synchronization.
Especially for the case of $Y=12m$ in the empty room, the big errors are at the percentage of $12\%$.
We can find the corresponding results at E8 and E16 shown in Figure \ref{fig:syncexp1}, where the successful detection rate is also much lower than other cases in synchronization.
Actually, since the successful detection rate in synchronization is above $80\%$ for most cases, the result would converge to the correct value if given enough time and the abnormal result would be eliminated.
Hence, we conclude that the ranging results are very good in these cases.


\subsection{Putting it All Together in a Severe Environment}
\label{subsec:all}
\begin{figure}[htpb]
    \vspace{-0.15in}
    \centering
    \includegraphics[width=3in]{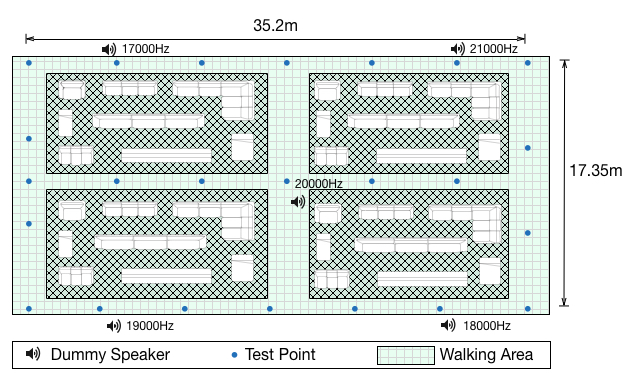}
    \caption{Map of the shopping mall.}
    \label{fig:map}
    \vspace{-0.1in}
\end{figure}

\begin{figure}[htpb]
    \vspace{-0.25in}
    \begin{subfigure}[b]{0.235\textwidth}
        \begin{center}
            \includegraphics[height=1.2in]{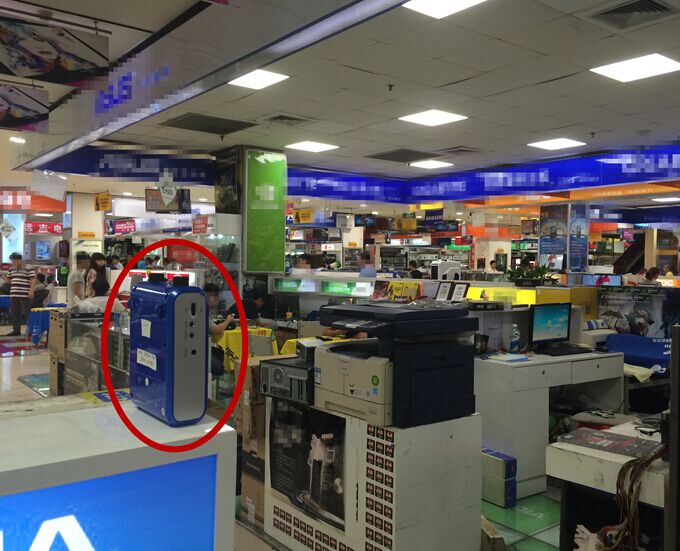}
        \end{center}
    \vspace{-0.1in}
        \caption{Shopping mall.}
        \label{fig:shoppingmall}
    \end{subfigure}
    \begin{subfigure}[b]{0.235\textwidth}
        \begin{center}
            \includegraphics[height=1.4in]{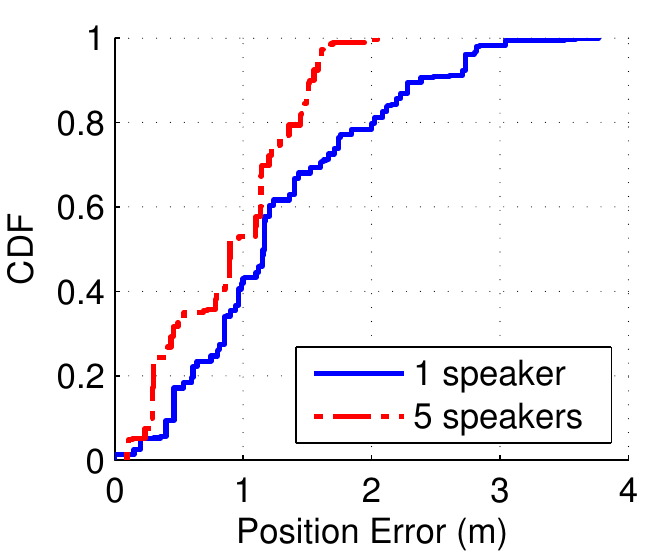}
        \end{center}
    \vspace{-0.15in}
        \caption{Position Errors}
        \label{fig:shoppingmallerror}
    \end{subfigure}
    \caption{(a) Shopping mall and the dummy speaker.  (b) Result of relative positioning (using 1 speaker), and absolute positioning (using 5 speakers).} 
    \label{fig:shopptingmallpic}
    \vspace{-0.1in}
\end{figure}

We evaluate \ourprotocol in a shopping mall,
where the environment is quite severe for acoustic based systems:
the shopping mall itself is broadcasting loud audios;
there are always people walking around who blocks the sight line of speakers or blocks the road that we have to turn walking direction.
Furthermore, as it may affect the business if we set up speakers on the ceiling and conduct frequent debugging (which may have better results), we only put the speakers at the side of the aisles, as shown in Figure \ref{fig:map}, \ref{fig:shoppingmall}.
Hence, our system has to deal with serious NLoS effects.

We evaluate the performance of positioning in two cases: a) \textit{relative} positioning by \textit{one speaker}. b) \textit{absolute} positioning by \textit{5 speakers} (like normal indoor localization). 
We choose a $35m \times 17m$ area (about $600 m^2$) in Figure \ref{fig:map}, and put 5 normal dummy speakersin this area.
Each speaker broadcasts signals at different central frequency, which are inaudible and not discovered by surrounding customers. 
We emulate the behavior of normal shopping users in evaluation: the experimenter stands at a test point and walks for a few steps (less than 6m) in a line; then he stops or turns the direction and continues walking, and so on. 
We gather 8 samples per point. 
Hence, we can evaluate the performance when leveraging all the walking segments to get the position.

We set central frequency of the speakers to 17000Hz, 18000Hz, 19000Hz, 20000Hz, 21000Hz, respectively.
The smart device differentiates the signals by using the subcomponent BPF in the Figure \ref{fig:overview}.
For example, if we need to analyze the signal of the second speaker (18000Hz), we set the frequency band of BPF which filters the signal at 18000Hz, and other signals are blocked.

The results show that these 5 speakers have much different performances in relative positioning, though they are the same product model.
The signal of speaker at 17000Hz only covers $13\%$ of the area, but the signal of speaker at 19000 and 20000 covers about $54\%$ and $51\%$ of the area. 
The reason of this diversity may be caused by several facts: anchor positions, quality of different anchor speakers, etc.
We leave the study on configuration of speakers in our future work.
Totally, the average coverage per speaker is $38\%$, which is about $222 m^2$ in our specific area.

We show the relative position errors when using one speaker in Figure \ref{fig:shoppingmallerror}.
Note that we only calculate the accuracy of the relative position where the starting point is covered by the signal of the speaker. 
Though we can still estimate position according to historical positioning result when there is no signal, we exclude the results of this case and obtains the direct result.
The results show that for one speaker, the position errors are under $1.2m, 2m$ at the percentage of $50\%$ and $80\%$.
The mean error of relative positioning is $1.28m$.

We also explore the localization capabilities when all 5 speakers are used as anchors. 
We evaluate the errors at all points and the results show that the position errors are under $1.5m$ at the percentage of $90\%$.
Since the average coverage per speaker is $38\%$, the smart device can receive audio from $38\%*5\approx 2$ speakers on average. 
The accuracy is intuitively better when using multiple signals for localization.

\subsection{Overhead}
The computation overhead is mainly caused by 3 components: displacement tracking (Including BPF, AGC, PLL), pulse detection and position estimation.
We run \ourprotocol using matlab R2013a on Mac OS, and the CPU is 3.1GHz Intel Core i5.
For 1 second of received samples, phase tracking, pulse detection, and position estimation takes 0.09s, 0.12s, 0.05s respectively. 
In fact, there is a trade-off between the overhead and accuracy. 
For example, we can use infinite BPF instead of finite BPF, which reduces the computation overhead significantly, but incurs larger errors.
For the smart devices, it is recommended to send the recorded samples to cloud server, and obtains the result from the cloud, which requires much less computation overhead, meanwhile with low energy consumption.


\section{Related Work}
\label{sec:relatedwork}
\subsection{Ranging}
There have been many localization systems which are based on ranging \cite{2013-MobiSys-Guoguoenablingfine,Priyantha:2000:CLS:345910.345917,6566822, DBLP:conf/mobicom/HarterHSWW99}.
They achieve considerable accuracy of ranging, but require special hardwares for synchronization purpose.
Specifically, the sender records sending time of signal which is used for ranging, while the receiver detects the arrival time of the signal.
Each individuals calculate the sending time or arrival time independently without referring any time information on other devices.
Hence, synchronization among devices is needed.
In Bat System \cite{DBLP:conf/mobicom/HarterHSWW99}, the base-station uses radio channel and communications for synchronization.
Cricket \cite{Priyantha:2000:CLS:345910.345917} uses special device to send the RF signal together with the ultrasound signal at the same time.
Then the receiver obtains the distance according to the different traveling time of the two signals.
Guoguo \cite{2013-MobiSys-Guoguoenablingfine} uses RF signals to synchronize all the acoustic anchors, 
the location can be obtained according to the differences of the receiving time by the phone. 
BeepBeep \cite{2007-SenSys-BeepBeephighaccuracy} calculates the distance between the phones.
It solves the synchronization problem by letting two phones emit acoustic signals and exchange the sending and receiving time via wireless channel.

\ourprotocol uses dummy speaker to implement synchronization and ranging.
The synchronization information is obtained by a novel position estimation method that it does not need any special hardwares or additional communication channels.
The other difference is that these systems are only based on ranging results of anchors which requires multiple speakers ($\ge 3$), while \ourprotocol also implements direction estimation from phone to speaker and only one speaker is needed for localization.



\subsection{Direction Estimation}
Most methods on direction estimation also require specialized hardwares, which use the directional antenna \cite{4711074,4509717,Niculescu:2004:VBS:1023720.1023727} or the antenna array  \cite{Joshi:2013:PLI:2482626.2482651,Xiong:2013:AFI:2482626.2482635, 4509717}.
For example, by rotating the beam of directional antenna, a receiver can pinpoint the direction of the AP as the direction that provides the highest received strength \cite{4509717}.
For the antenna array \cite{Joshi:2013:PLI:2482626.2482651,Xiong:2013:AFI:2482626.2482635, 4509717}, the receiving time of the signal by each antenna is different, and magnitude of the difference corresponds to angle of the arrival signal.
 
There have been proposals without requirement of specialized hardwares as well.
 \cite{2011-MOBICOM-Iamantenna} emulates the functionality of a directional antenna by rotating the phone around the user's body, to locate outdoor APs.
\cite{2011-SenSys-feasibilityrealtime} leverages multiple microphones of the smartphone and communication channels for positioning within 4 meters, which is used for short-distance positioning and phone-to-phone games.
Some other methods leverage Doppler effects by swinging \cite{2012-MobiQuitous2011-ProposalDirectionEstimation} or shaking \cite{DBLP:journals/corr/HuangXLLMYL13} the phone.
\cite{DBLP:conf/mobicom/ZhangLHLZJFJL14} calculates direction by head nodding or shaking using smart glasses.
They are based on different frequency shift when the phone are moving at different directions.
Compares to  \cite{2012-MobiQuitous2011-ProposalDirectionEstimation,DBLP:journals/corr/HuangXLLMYL13}, \ourprotocol makes further steps that a user can obtain direction without any additional actions on the phone so that s/he can get the real-time direction while walking.
Furthermore, \cite{DBLP:journals/corr/HuangXLLMYL13} requires only the speakers as anchors as well, but does not address the ranging problem, while \ourprotocol can compute both the direction and distance from the phone to the speaker.

\section{Conclusion}
\label{sec:conclusion}
We propose and implement \ourprotocol, a localization scheme that calculates the relative position from a smart device to a dummy speaker.
The dummy speaker only needs to emit acoustic signals at non-audible frequency, so that COTS speakers can serve as anchors.
Furthermore, \ourprotocol directly obtains both distance and direction from smart device to speaker, which is quite different from existing localization systems that are capable of obtaining only the distance or the direction.
As a result, \ourprotocol only requires one anchor for localization, while others need multiple anchors for calculating the final position, such as trilateration.
By pushing the limit of the anchor's number, \ourprotocol is not only capable of indoor localization, but also has a potential for wider applications, such as augmented-reality or mobile social applications. 


{\small
\vspace{0.05in}
\bibliographystyle{acm}

\bibliography{bib}
}

\end{document}